\documentclass[a4paper, intlimits, 11pt,reqno]{amsart}

\usepackage[hmargin=1.4in,vmargin=1.2in]{geometry} 
\usepackage{enumitem}
\usepackage{mathtools,dsfont}
\usepackage[utf8]{inputenc}
\usepackage{float}
\usepackage{times}
\usepackage[english]{babel}
\usepackage{graphicx}
\usepackage{caption}
\usepackage[labelformat=simple, labelfont=rm ]{subcaption}
\usepackage{bm}

\usepackage{booktabs}
\usepackage{pdflscape}
\usepackage{rotating}
\usepackage{array}       
\usepackage{adjustbox}   
\usepackage{eurosym}
\usepackage{hyperref}
\usepackage{natbib}
\usepackage[normalem]{ulem}
\usepackage{multirow}
\usepackage{upgreek}
\usepackage{amsthm, amssymb, amsfonts}
\usepackage{colortbl}
\usepackage{accents}
\usepackage{url}
\usepackage{comment}

\newcolumntype{C}[1]{>{\centering\arraybackslash}p{#1}}

\def \proof{\noindent \emph{\textbf{Proof.} $\, $}} 
\def \finproof{\ensuremath{\square}}

\newcommand {\norm}[1]{\left\lVert#1 \right\rVert} 

\DeclareMathOperator*{\argmin}{arg\,min}

\newtheorem{theorem}{Theorem}[section]
\newtheorem{corollary}[theorem]{Corollary}

\newtheorem{lemma}[theorem]{Lemma}
\newtheorem{proposition}[theorem]{Proposition}

\theoremstyle{definition}
\newtheorem{definition}{Definition}[section]
\newtheorem{assumption}[definition]{Assumption}

\theoremstyle{remark}
\newtheorem{remark}{Remark}[section] 

\newtheorem{notation}[remark]{Notation}

\numberwithin{equation}{section}

\hypersetup {
	colorlinks  = true, %
	urlcolor   = black, %
	linkcolor  = black, %
	citecolor  = black %
}

\makeatletter
\renewcommand\@makefnmark{ \hbox{\@textsuperscript{\normalfont\color{blue}\@thefnmark}} }
\renewcommand\@makefntext[1]{%
	\parindent 1em\noindent
	\hb@xt@1.8em{%
		\hss\@textsuperscript{\normalfont\@thefnmark}}#1}
\makeatother

\makeatletter
\newcommand\@notni[2]{\mathrel{\rotatebox[y=#1]{180}{$#2\notin$}}}
\newcommand\notni{
	\mathchoice
	{\@notni{0.57ex}\displaystyle}
	{\@notni{0.57ex}\textstyle}
	{\@notni{0.39ex}\scriptstyle}
	{\@notni{0.26ex}\scriptscriptstyle}
}
\makeatother

\makeatletter
\renewcommand\p@subfigure{\thefigure\,}
\makeatother

\begin{document}

\title{Sensitivity Analysis of emissions Markets: A Discrete-Time Radner Equilibrium Approach}

\author{ St\'ephane Cr\'epey\textsuperscript{\lowercase {a}},
         Mekonnen Tadese\textsuperscript{\lowercase {b, c}},
         Gauthier Vermandel\textsuperscript{\lowercase {d}} \\ 
         \today
        }
	
\date{  {\footnotesize{}{}{}{} This version: \today}     }

\begin{abstract}
	Emissions markets play a vital role in emissions reduction by incentivizing firms to minimize costs. However, their effectiveness heavily depends on the decisions of policymakers, future economic activity, and the availability of abatement technologies. 
	This study investigates how variations in regulatory standards, firms' abatement costs, and emission levels influence allowance prices and firms' abatement efforts.  
	This is done in a Radner equilibrium framework that incorporates intertemporal decision-making and uncertainty, enabling a comprehensive analysis of market dynamics and outcomes. 
	The findings provide valuable insights for policymakers aiming to enhance the design and efficiency of emissions trading systems through a deeper understanding of stakeholder responses across varying market conditions.

    \vspace{5pt}
    
    \vspace{5pt}
\end{abstract}

{ \let\thefootnote\relax\footnotetext{
{This research has benefited from the support of Chair Capital Markets Tomorrow: Modeling and Computational Issues under the aegis of the Institute Europlace de Finance,  a joint initiative of Laboratoire de Probabilit\'es, Statistique et Modélisation (LPSM) / Université Paris Cit\'e and  Cr\'edit Agricole CIB,  and of Chair Stress Test, Risk Management and Financial Steering, led by the French Ecole Polytechnique and its foundation and sponsored by BNP Paribas.}
}}
{ \let\thefootnote\relax\footnotetext{\textsuperscript{a} \textit{Laboratoire de Probabilités, Statistique et Modélisation (LPSM), Sorbonne Université et Université Paris Cit\'e, CNRS UMR 8001.} stephane.crepey@lpsm.paris.}
}

{ \let\thefootnote\relax\footnotetext{\textsuperscript{b} \textit{Laboratoire de Probabilités, Statistique et Modélisation (LPSM), Sorbonne Université.}  demeke@lpsm.paris. 
The research of M. Tadese is co-funded by a  grant from the program PAUSE, Collège de France.}
}
{ \let\thefootnote\relax\footnotetext{\textsuperscript{c} \textit{ Department of Mathematics, Woldia University, Ethiopia.}
}

{\let\thefootnote\relax\footnotetext{\textsuperscript{d} \textit{Centre de Math\'ematiques Appliqu\'ees (CMAP), Ecole Polytechnique, France,} / \textit{Universit\'es Paris-Dauphine \& PSL, France} / \textit{Banque de France, DECAMS, France.} gauthier@vermandel.fr.}
}
 
\maketitle

\section{Introduction}

\subsection{Motivation, Approach, and Theoretical Contribution}

The global effort to combat climate change, exemplified by the Paris Agreement \citep{united2015}, has driven the adoption of regulatory measures aimed at reducing greenhouse gas emissions.
Market-based mechanisms, particularly emissions trading systems (ETS), have emerged as key policy tools \citep{Worldbank2025}.
In Europe, carbon emissions are regulated through the EU ETS, which operates on a ``cap-and-trade'' principle \citep{woerdman2015eu}.
Under this system, a cap is imposed on the total emissions of each polluter, and emission rights are tradable in the market, thereby enhancing overall cost-effectiveness.
Its popularity has grown, with more than 37 cap-and-trade systems (covering 23\% of global emissions) currently active worldwide \citep*{blanchard2022fighting, Worldbank2025, ICAP2025}.

Despite their popularity, emission markets are complex systems influenced  by various factors, including regulatory policies, market behaviors, and economic conditions. 
It is essential for regulators to develop a deeper understanding of the latter to maintain the effectiveness and efficiency of the policies.
Polluters should also comprehend these mechanisms to effectively manage carbon pricing risks.
One effective way to analyze such markets is through sensitivity analysis, which constitutes the main focus of this study.
Sensitivity analysis assesses how variations in input parameters affect key market outcomes---such as allowance prices and firms' abatement efforts---thereby providing insights into both the direction and magnitude of these effects.

Prior studies have primarily focused on assessing the impact of individual risk factors on market outcomes \citep*[see, for instance][]{ Seifert2008, carmona2009, guo2019, huang2022}. 
While these analyses provide a general understanding of how these factors affect market outcomes, two key gaps remain unaddressed.
First, there is a lack of systematic comparison of the relative magnitudes of the effects exerted by different risk factors.
In particular, the extent to which various factors interact to produce mutually amplifying or counteracting effects remains insufficiently investigated.
Second, much of the existing literature relies on representative-agent models, which capture aggregate trends but overlook interactions among individual firms. 
Moreover, sensitivity analyses have typically concentrated on allowance prices, with limited attention to firm-level abatement responses.

To address these shortcomings, this paper uses elasticity-based sensitivity analysis to quantify the magnitude and interactions of key drivers. 
We also explicitly account for differences between individual firms and their interactions within the market.
To the best of our knowledge, such an elasticity analysis provides a novel contribution to the sensitivity analysis of emissions markets.

To achieve our objective, we employ a discrete-time Radner equilibrium cap-and-trade model, similar to \citet*{carmona2009}, \citet*{carmona2010market} and \citet{aid2023}. 
The Radner equilibrium framework is particularly well-suited to incomplete markets, as it captures uncertainty, limited availability of financial instruments, and firms' expectations---offering a more realistic depiction of market dynamics and inefficiencies than complete-market models. 
Within this setting, our analysis focuses on two primary sources of uncertainty in the economy: firms' emission levels and the emissions cap, the latter defined as a fraction of stochastic baseline emissions.

\subsection{Review of the Literature}
A substantial body of research has investigated the determinants of carbon allowance prices, emphasizing how regulatory policies and economic conditions influence market outcomes.
\citet*{Seifert2008} investigate the dynamics of allowance prices and analyze key characteristics of the EU ETS.
Their study provides both qualitative and quantitative insights into how changes in the penalty rate, emissions cap, marginal abatement cost, and the mean and volatility of emissions affect allowance prices.
In addition, they examine how the volatility of the allowance price responds to variations in the penalty rate, the marginal abatement cost, and the volatility of emissions.
\citet*{carmona2009} examine the impact of stochastic abatement costs on allowance prices, demonstrating that the mean cost parameter is the primary price driver, with volatility playing a comparatively smaller role.
Similarly, \citet*{guo2019} analyze the impact of uncertainties in abatement costs and emission rates, focusing primarily on their implications for the abatement planning process.
\citet*{hitzemann2018} study the distinct properties of allowance prices across multiple compliance periods.
In a broader macro-environmental context, \citet*{huang2022}  develop an equilibrium model for allowance price based on population growth and climate change, examining how allowance prices are affected by GDP growth and global warming. 
More recently, \citet[Section 5.2]{aid2023}  propose a continuous-time model structurally similar to ours but provide limited mathematical detail and minimal numerical implementation.

Some studies adopt simulation-based or empirically calibrated approaches to evaluate policy impacts. 
\cite{jiang2021} develop a simulation-based dynamic assessment model for carbon markets that replicates allowance allocation, emissions, trading behavior, and price formation under empirical calibration. 
Their primary focus is on evaluating the impact of emission reduction targets on carbon prices.
Using partial equilibrium approach, \cite{Yu2020} also study the interaction between two carbon markets: an emission allowance market and a carbon offset market, whereby firms purchase credits representing verified emissions reductions achieved outside the regulated allowance system.
They also study the sensitivity of  allowance prices with respect to penalty rate and quadratic abatement cost coefficient, emissions cap and other parameters related to the second offset market, in a representative-agent framework. 

Most sensitivity analyses in these work are based on a representative agent framework, which neglects the strategic interactions among individual firms within the market. 
Furthermore, they largely omit elasticity analysis, an essential tool for comparing the relative impacts of different parameters on allowance prices and each firm's abatement efforts.
Whenever available, sensitivity analysis in this context typically focuses on allowance prices and devotes limited attention to individual firms' abatement efforts. 

\subsection{Main Results} 
This article delivers three key results. 
The first one discusses the impact of regulatory standards on carbon pricing. 
When firms face strict mitigation policies that include significant penalties for exceeding emissions caps, the average prices of allowances increase. 
Consequently, firms become more willing to reduce carbon emissions.
The result that higher penalties reduce emissions is consistent with cost-benefit analyses in integrated assessment models \citep[e.g.][]{nordhaus2024}, where a higher carbon tax strengthens abatement incentives. In our cap-and-trade setting, the penalty plays an analogous role, effectively acting as a tax in states of nature where aggregate emissions exceed the cap. 
Our results also indicate that higher penalties not only increase the mean of allowance prices, but also increase  price volatility. 
This suggests that while penalties can effectively drive firms to reduce emissions, they also introduce risk into the carbon market. 
In contrast, we observe that tighter emissions caps exert a stronger and more consistent upward pressure on prices and have comparatively stabilizing effects on volatility. 
From a policy standpoint, cap adjustments provide a more reliable means of stabilizing prices, while reliance on penalties constitutes a comparatively more volatile approach.

The second key result examines how abatement costs affect firms' decisions to reduce carbon emissions. 
Firms with higher abatement costs tend to be more responsive in adjusting their emission reduction efforts when their costs fluctuate. 
As a result, a rise in abatement costs for firms that initially face lower costs can, somewhat counterintuitively, lead to an overall increase in economy-wide abatement efforts.
Indeed, as the difference in abatement costs between firms widens, firms will increasingly rely on the ETS market to trade allowances, exploiting the lower-cost abatement opportunities available to other firms. 
Additionally, we found that an increase in abatement costs leads to lower volatility of allowance prices in line with \citet{hitzemann2018}---an inverse relationship commonly referred to as the 'leverage effect' in economic literature.

The final key result addresses the impact of uncertainty in firms' emissions levels. 
In our model, firms face stochastic variations in their emissions, which also introduces uncertainty about the amount of allowances they need to purchase to comply with the cap.
We find that the expected value of firms' business-as-usual (BAU) emissions is a primary factor driving allowance prices and abatement efforts. 
Variability factors such as the standard deviation of firms' BAU emissions and the correlations between emissions across firms, instead, have minimal influence on both allowance prices and abatement strategies.

\subsection{Outline of the Paper and Standing Notation}
The remainder of the paper is structured as follows.
Section~\ref{sec:model} introduces the carbon pricing model and establishes the existence and uniqueness of the corresponding Radner equilibrium.
Section~\ref{sec:senstivity} specializes the model to a Gaussian setting and derives explicit expressions for the allowance price and its standard deviation as functions of the optimal abatement plans and other model parameters.
Section~\ref{sec:theoretical_sensitivity} presents theoretical sensitivity analysis results using the implicit function theorem.
Section~\ref{sec:numerical} provides a numerical evaluation of the model and analyzes the sensitivity of firm-level abatement efforts and allowance prices to specific parameters.
Section~\ref{sec:conclusion} concludes.
Auxiliary results related to Sections~\ref{sec:model} and \ref{sec:theoretical_sensitivity} are provided in Sections~\ref{app:proof3} and \ref{app:equal_efoort}. \\
  
Let $0\,..\,T$, for some positive integer $T$, represent an emission allowances compliance period, that is, a future time interval at the end of which market emissions are released and penalties may apply \citep{carmona2009, carmona2010market}.
Let a filtered probability space $(\Omega, (\mathfrak{F}_t)_{t=0}^T, \mathbb{P})$ with $\mathfrak{F}_0 = \{\emptyset, \Omega\}$ represent the views of the users on the future. 
By default, we omit any dependence on $\omega\in \Omega$ in the notation. 
Expectation and correlation are denoted by $\mathbb{E}$ and $\mathrm{corr} $, respectively, while $\mathbb{E}_t$ and $\mathbb{P}_t$ denote the conditional expectation and conditional probability given the sigma-algebra $\mathfrak{F}_t$.
For $p =1, 2, \infty$, we define $L^p_t = \{(X_s)_{s=0}^t \colon  X_s \mbox{ is }  \mathfrak{F}_s  \mbox{ measurable and $p$ integrable for all } 0\leq s\leq t\}$, where $p$-integrability with $p = \infty$ refers to essential boundedness.
Inequalities between random variables are understood almost surely. 
The Euclidean norm of a vector $x \in \mathbb{R}^d$ is denoted by $\norm{x}$.
	
\section{General Radner Equilibrium Setup}\label{sec:model}

We consider a set of $n$ firms, indexed by $i$. 
Let $e^i = (e^i_t)_{t=0}^{T-1} \in L^2_{T-1}$ denote the emissions of firm $i$ under business-as-usual (BAU) case, where $e^i_t$ is its emissions at time $t$.
The total emissions of firm $i$ and the aggregate emissions in the economy at time $T$ under the BAU case are then given, respectively, by
\begin{equation*}
	\sum_{t=0}^{T-1} e^i_t  
	\qquad  	\text{and} 	\qquad  
	\sum_i \sum_{t=0}^{T-1} e^i_t  . 
\end{equation*}

Let $A_T$ denote the total emissions cap for the economy over the entire compliance period. 
Each firm $i$ receives an allocation $A^i_T$ of the total cap. 
These allowances are granted free of charge and can  be distributed dynamically over time. 
Let $A = (A_t)_{t=0}^{T}$ in $L^2_T$ denote the cumulative number of allowances distributed up to time $t$, and let $A^i = (A^i_t)_{t=0}^{T}$ represent the cumulative allowances allocated to firm $i$.
If the cumulative emissions $\sum_{t=0}^{T-1} e^i_t$ of firm $i$ exceed its allowable cap $A^i_T$ at the end of the compliance period, the firm incurs a penalty of $\lambda$ euros per unit of excess emissions, where $\lambda > 0$ is a constant.

Each firm $i$ can reduce its emissions $e^i_t$ at time $t \in 0\, .. \, T-1$ by a proportion $\alpha^i(t) \in [0,1]$, resulting in an abatement level of $\alpha^i(t) e^i_t$ at time $t$. 
Consistently with a convex abatement cost curve as per \cite{nordhaus2024}, the associated cost of such an abatements can be given by
\begin{equation}\label{eq:abatecost}
	\sum_{t=0}^{T-1} \left[ \mathrm{k}^i \alpha^i (t) e^i_t  +\tfrac{\gamma^i}{2}\big(\alpha^i (t) e^i_t \big)^2 \right],
\end{equation}
where $\mathrm{k}^i$ and $\gamma^i$ are positive constants, while the abated proportion $\alpha^i(t) \in [0, 1]$ is a control of firm $i$ at time $t$. 
Given an abatement plan  $\alpha = (\alpha^1, \dots, \alpha^{ n })^\top$ in $\mathcal{A}  =[0,1]^{ n T}$ for all firms, where each $\alpha^i=\big(\alpha^i (0), \dots, \alpha^i (T-1) \big)$ in $\mathcal{A}^i=[0,1]^T$, the cumulative emissions of firm  $i$ and the aggregate emissions at the end of the compliance period after abatement become
\begin{equation}\label{eq:agregate_emision}
    E^{i,\alpha^i}_T   = \sum_{t=0}^{T-1} \big(1-\alpha^i (t) \big) e^i_t 
	\quad 
	 \text{and} 
	\quad  E^\alpha_T    =  \sum_{i } E^{i,\alpha^i}_T. 
\end{equation}

Consistent with the structure of carbon market, firms can trade their allowances at any time $t \in 0\ ..\ T$. 
Trading serves two primary purposes: first, it allows firms to manage and smooth their emissions over time without relying exclusively on abatement; second, it enables firms to exploit differences in abatement costs by trading allowances with other firms, thereby optimizing their overall compliance strategies.
Let $P=(P_t)_{t=0}^{T} \in L^\infty_T$,  where the ``allowance price" $P_t$  represents the price at which a buyer of a forward allowance contract issued at time $t$ agrees to purchase the allowance at time $T$.
Let  $Q^i=(Q^i_t)_{t=0}^{T} \in L^1_T$, where $Q^i_t$ represents the number of forward  allowances held by firm $i$  at time $t$.
For firm $i$, the  trading cost of the strategy $Q^i$ (with  $Q^i_{-1}=0$) is  
\begin{equation}\label{eq:tradcost}
    \sum_{t=0}^T (Q^i_t-Q^i_{t-1})P_t=  - \sum_{t=0}^{T-1} Q^i_t \Delta P_t + Q^i_T P_T,
\end{equation}	
where $\Delta P_t =P_{t+1}-P_t,\; t\in 0\, ..\, T-1$.
The terminal penalty cost of firm $i$ is equal to  
\begin{equation}\label{eq:tercost}
	\lambda\big(E^{i,\alpha^i}_T  -A^i_T-Q^i_T\big)^+.
\end{equation}
The ensuing loss profile of firm $i$ is the sum of~\eqref{eq:abatecost},~\eqref{eq:tradcost} and~\eqref{eq:tercost}, i.e.
\begin{equation}\label{eq:lossprofilefree}
	C^{i, P}  ( \alpha^i, Q^i )  = \sum_{t=0}^{T-1} \left[ \mathrm{k}^i \alpha^i (t) e^i_t +\tfrac{\gamma^i}{2} \big(\alpha^i (t) e^i_t \big)^2 - Q^i_t \Delta P_t\right] + Q^i_T P_T + \lambda\big(E^{i,\alpha^i}_T -A^i_T-Q^i_T\big)^+.
\end{equation}
The price $P$, the abatement plan $\alpha^i$ and  the trading  strategy $Q^i$ are determined by a Radner equilibrium as per Definition~\ref{def:equilibrium}, where Radner equilibrium quantities are represented by bold letters.  
A recapitulation of the  notation is provided in Table~\ref{tab:not}.
\begin{table}[htp]  
	\caption{Main notation. Deterministic and  random  time dependencies are denoted by $(t)$ and $\cdot_t$. 
   } 
		\label{tab:not}
	\begin{centering}
		{
			\begin{tabular}{@{}cll@{}} 
				\toprule
				\multicolumn{3}{c}{\emph{Regulatory standard} }\\
				\cmidrule{1-3}
				
				$T$ & & end of the compliance period\\
				$A^i_t$ && firm $i$’s granted endowment of emissions allowances until time $t \in 0\,..\,T$ \\
				$\lambda$ & & penalty per ton of excess emissions at time $T$\\
				\cmidrule{1-3}
				
				\multicolumn{3}{c}{\emph{Emissions under the business-as-usual (BAU) case } }\\
				\cmidrule{1-3}
				$e^i_t$  &    & emissions of firm $i$ under the BAU case at time $t\in 0\,..\,T-1$  \\
				\cmidrule{1-3}
				
				\multicolumn{3}{c}{\emph{Abatement plans and  emissions after abatement} }\\
				\cmidrule{1-3}
				$\alpha^i (t)$ & & fractional abatement plan of firm $i$ at time $t\in 0\,..\,T-1$\\
				$E^{i,\alpha^i}_T$ & & cumulative emissions (after abatement) of firm $i$ at time $T$\\
				$\mathrm{k}^i$ && linear abatement cost  coefficient  of firm $i$\\
				$\gamma^i$ && quadratic abatement cost  coefficient of firm $i$\\
				\cmidrule{1-3}
				
				\multicolumn{3}{c}{\emph{Trading strategies and allowance price} }\\
				\cmidrule{1-3}
				$P_t$ & & price of emissions allowances at time $t \in 0\,..\,T$\\
				$Q^i_t$ & & quantity of emission allowances traded by firm $i$ at time $t \in 0\,..\,T$\\
				\bottomrule
				
			\end{tabular}
		}
	\end{centering}
\end{table}

\begin{definition}\label{def:equilibrium}
	The triple  $\left( (\pmb{\alpha}^i)_i, (\mathbf{Q}^i)_{i}, \mathbf{P} \right) \in  \mathcal{A}\times (L^1_T)^n \times L^\infty_T$ forms a Radner equilibrium,  where $\mathbf{P}$ is the forward allowances price, $\pmb{\alpha}^i$ and  $\mathbf{Q}^i$ are the abatement plan  and trading strategies of firm $i$, such that
	{ \rm\hfill\break $\bullet$ (optimality condition relative to each market participant $i$) }
	\begin{equation*}
		\mathbb{E} \big[C^{i,\mathbf{P} }  ( \pmb{\alpha}^i, \mathbf{Q}^i )\big] \leq \mathbb{E} \big[C^{i,\mathbf{P} }  ( \alpha^i, Q^i )\big], \; \alpha^i \in \mathcal{A}^i, Q^i \in L^1_T; 
	\end{equation*}
	{$\bullet$  (zero clearing condition)}
	\begin{equation*}
		\sum_{i } \mathbf{Q}^i_t = 0, \quad t\in 0\, ..\,T.
	\end{equation*}
\end{definition}

\begin{assumption}\label{ass:uniqueness}
	The random variable $E^\alpha_T -A_T $ has continuous distribution for each  $\alpha\in \mathcal{A}$.
\end{assumption}
Let $\alpha$ in $\mathcal{A}$ be an abatement plan of all firms.
For ease of notation, we introduce  
\begin{equation}\label{eq:terminalprice}
	\xi^\alpha  = \lambda\; \mathds{1}_{\{ E^\alpha_T- A_T > 0\}}.
\end{equation}

\begin{lemma}\label{lem:terminalstrat}
If $P$ is the martingale closed by $\pmb{\xi}^{\pmb{\alpha} }$, i.e. $P_t = \mathbb{E}_t[\pmb{\xi}^{\pmb{\alpha} }]$ for $t\leq T$, then  for each $\alpha \in \mathcal{A}$ and firm $i$,
\begin{equation*}
    \min_{Q^i \in L^1_T } \mathbb{E}[C^{i, P}(\alpha^i, Q^i)] =  \mathbb{E} \Big [\sum_{t=0}^{T-1}  k^i \alpha^i (t) e^i_t +\tfrac{\gamma^i}{2} \big(\alpha^i (t) e^i_t \big)^2 + \pmb{\xi}^{\pmb{\alpha}} \big(E^{i,\alpha^i}_T -A^i_T \big) \Big].
\end{equation*}
\end{lemma}

\proof
Let $P$ is the martingale closed by $\pmb{\xi}^{\pmb{\alpha} }$.
It suffices to show 
\begin{equation}\label{eq:thelema}
		\pmb{\xi}^{\pmb{\alpha}} (E^{i,\alpha^i}_T - A^i_T)  = \min_{ Q^i  \in L^1_T } \big\{ Q^i_T \pmb{\xi}^{\pmb{\alpha}} + \lambda  \big(E^{i, \alpha^i}_T -A^i_T-Q^i_T\big)^+\big\}. 
\end{equation}
$Q^i\in L^1_T$ is a minimizer for~\eqref{eq:thelema} if  $Q^i_T(\omega)$ coincides for almost every $\omega$ with some minimizer $x^{i, \alpha^i, \pmb{\alpha} } (\omega)$  of 
\begin{equation*}
	\mathbb{R}  \ni	x \stackrel{f^{i, \alpha^i, \pmb{\alpha}}_\omega }{\mapsto} x \pmb{\xi}^{\pmb{\alpha}}(\omega) + \lambda \big( E^{i,\alpha^i}_T(\omega) -A^i_T(\omega)- x \big)^+.
\end{equation*}
For  $\omega\in \{\pmb{\xi}^{\pmb{\alpha}} =0\}$, $f^{i, \alpha^i, \pmb{\alpha}}_\omega$ reduces to
\begin{equation*}
	f^{i, \alpha^i, \pmb{\alpha}}_\omega(x)= \lambda \big( E^{i,\alpha^i}_T(\omega) -A^i_T(\omega)-x \big)^+, 
\end{equation*}
with minimum  equal to $0$, reached for any $x \geq  E^{i,\alpha^i}_T(\omega)-A^i_T(\omega) $.
If instead  $\omega \in \{\pmb{\xi}^{\pmb{\alpha}} =\lambda\}$, then  
\begin{equation*}
	f^{i, \alpha^i, \pmb{\alpha}}_\omega(x)   =  \lambda x + \lambda \big( E^{i,\alpha^i}_T(\omega)-A^i_T(\omega) -x \big)^+,
\end{equation*}
with  minimum  equal to $\lambda \big(E^{i,\alpha^i}_T(\omega)-A^i_T(\omega)\big)$, reached for any $x \leq E^{i,\alpha^i}_T(\omega)-A^i_T(\omega)$.
This ensures~\eqref{eq:thelema}.
As $P$ is a martingale closed by $\pmb{\xi}^{\pmb{\alpha}}$ and $Q^i$ is adapted, we have $\mathbb{E}[\sum_{t=0}^{T-1} Q^i_t \Delta P_t] =0$, $Q^i \in L^1_T$.
Hence, the Lemma follows from~\eqref{eq:thelema}.\ \finproof\\

 Proposition~\ref{prop:Fallocation}  is a variant of \citet[Theorem 1]{carmona2009} and \citet[Proposition 3.1]{carmona2010market}, with similar proof, provided for the reader's convenience, in Section~\ref{app:proof3}. 

\begin{proposition}\label{prop:Fallocation}
    If a price  process $\mathbf{P}$, an abatement plan $\pmb{\alpha}$ and trading strategies  $(\mathbf{Q}^i)_{i }$  constitute a Radner equilibrium, then  $\mathbf{P}$ is the martingale closed by  $\pmb{\xi}^{\pmb{\alpha}} $.    
    The optimal expected cost of firm $i$
    \begin{equation*}
		\mathbb{E} \big [C^{i,\mathbf{P} }  (\pmb{\alpha}^i, \mathbf{Q}^i ) \big]= \mathbb{E} \Big [\sum_{t=0}^{T-1}  \left( \mathrm{k}^i \pmb{\alpha}^i (t) e^i_t +\tfrac{\gamma^i}{2} \big(\pmb{\alpha}^i (t) e^i_t \big)^2 \right) + \pmb{\xi}^{\pmb{\alpha}} \big(E^{i,\pmb{\alpha}^i}_T -A^i_T \big) \Big].
    \end{equation*} 
\end{proposition}
\proof See Section~\ref{app:proof3}.

Let $\displaystyle R(\alpha)= \mathbb{E} \Big[ \sum_{i} \sum_{t=0}^{T-1} \Big(\mathrm{k}^i \alpha^i (t) e^i_t +\tfrac{\gamma^i}{2} \big(\alpha^i (t) e^i_t \big)^2 \Big)  + \xi^\alpha (E^{\alpha}_T -A_T) \Big]$. 

\begin{theorem}\label{prop:optimaleffort1}
	An abatement plan	$\pmb{\alpha} \in \mathcal{A}$ is optimal if and only if 
	\begin{equation}\label{eq:socialyoptimal}
		R(\pmb{\alpha}) = \min_{ \alpha \in \mathcal{A} } R(\alpha).
	\end{equation}
	Moreover, there exists a unique optimal abatement plan and a unique  allowance price.
\end{theorem}

\proof
Let $\big(\pmb{\alpha}, (\mathbf{Q}^i)_i, \mathbf{P} \big)$ be a Radner  equilibrium.
Then, by Proposition~\ref{prop:Fallocation}, the price $\mathbf{P}$ is the martingale closed by $\pmb{\xi}^{\pmb{\alpha}}$.
Now, for any $\alpha \in \mathcal{A}$, define the trading strategy $Q^{i, \alpha}$ by  $Q^{i, \alpha}_t =0$, $t=0\,..\,T-1$, and $Q^{i, \alpha}_T=E^{i, \alpha^i}_T-A^i_T -(E^\alpha_T -A_T)/n$.
Proposition~\ref{prop:Fallocation}, combined with the optimality condition for each firm, implies
\begin{align*}
	R(\pmb{\alpha}) & = \sum_i \mathbb{E} \big[C^{i,\mathbf{P} }  (\pmb{\alpha}^i, \mathbf{Q}^i )\big] \leq \sum_{i} \mathbb{E} \big[C^{i,\mathbf{P} }  (\alpha^i, Q^{i, \alpha} )\big].
\end{align*}
As $\mathbf{P}$ is a martingale and  $Q^{i, \alpha}$ is adapted, substituting  $Q^{i, \alpha}$ for $Q^i$ in~\eqref{eq:lossprofilefree} yields 
\begin{align*}
	\sum_{i  }\mathbb{E} \big[C^{i,\mathbf{P} }  (\alpha^i, Q^{i, \alpha} )\big] 
	& = \mathbb{E} \bigg[\sum_{i} \sum_{t=0}^{T-1} \Big[ \mathrm{k}^i \alpha^i (t) e^i_t +\tfrac{\gamma^i}{2} \Big(\alpha^i (t) e^i_t \Big)^2 \Big]  + \lambda  (E^{\alpha}_T -A_T)^+\bigg] \\
	& = R(\alpha). 
\end{align*}
It follows that $R(\pmb{\alpha}) \leq R(\alpha)$, $\alpha \in \mathcal{A}$, i.e.~\eqref{eq:socialyoptimal} holds.

\noindent 
Conversely, suppose that the abatement plan $\pmb{\alpha}$ satisfies~\eqref{eq:socialyoptimal}.
We need to show the existence of the price $\mathbf{P} \in L^\infty_T$ and of trading strategies $\mathbf{Q}^i \in L^1_T$ such that the triple $( \pmb{\alpha}, (\mathbf{Q}^i)_i, \mathbf{P})$ constitutes a Radner equilibrium.
Define the trading strategy $\mathbf{Q}^i$ by  $\mathbf{Q}^i_t = 0$, $t < T$ and  $\mathbf{Q}^i_T = E^{i,\pmb{\alpha}^i }_T -A^i_T -(E^{\pmb{\alpha}}_T -A_T)/n$, and  let $\mathbf{P}$ denote the martingale closed by $\pmb{\xi}^{\pmb{\alpha}}$. 
By construction, the strategies $\mathbf{Q}^i$ satisfy the market-clearing condition.
It remains to verify the individual optimality condition for each firm $i$. 
In view of Lemma~\ref{lem:terminalstrat}, the optimality condition of firm $i$ is equivalent to
\begin{align*}
    &\mathbb{E} \Big [\sum_{t=0}^{T-1} \Big(  \mathrm{k}^i \pmb{\alpha}^i (t) e^i_t +\tfrac{\gamma^i}{2} \big(\pmb{\alpha}^i (t) e^i_t \big)^2 \Big) + \pmb{\xi}^{\pmb{\alpha}} \big(E^{i,\pmb{\alpha}^i}_T -A^i_T \big) \Big] \notag \\
      &  \qquad \leq \; \mathbb{E} \Big [\sum_{t=0}^{T-1} \Big( \mathrm{k}^i \alpha^i (t) e^i_t + \tfrac{\gamma^i}{2} \big(\alpha^i (t) e^i_t \big)^2 \Big) + \pmb{\xi}^{\pmb{\alpha}} \big(E^{i,\alpha^i}_T -A^i_T \big) \Big], \;  \alpha^i \in \mathcal{A}^i.
\end{align*}
Since $\pmb{\xi}^{\pmb{\alpha}} \big(E^{i,\alpha^i}_T -A^i_T \big) = \pmb{\xi}^{\pmb{\alpha}} \left( \sum_i \sum_{t=0}^{T-1}e^i_t - A^i_T \right) -\sum_{t=0}^{T-1} \pmb{\xi}^{\pmb{\alpha}}  \alpha^i(t) e^i_t $, for any $\alpha^i\in \mathcal{A}^i$, the inequality above is equivalently expressed as
\begin{equation*}
	\pmb{\alpha}^i  \in \argmin_{\alpha^i \in \mathcal{A}^i } \mathbb{E} \Big [\sum_{t=0}^{T-1} \Big( \mathrm{k}^i \alpha^i (t) e^i_t +\tfrac{\gamma^i}{2} \big(\alpha^i (t) e^i_t \big)^2  - \pmb{\xi}^{\pmb{\alpha}} \alpha^i(t) e^i_t \Big) \Big].
\end{equation*}
Therefore, the optimality condition for firm $i$  is equivalent to the condition that,  for $t\in 0\,..\,T-1$, 
\begin{equation*}
	\pmb{\alpha}^i(t)  \in \argmin_{x \in [0,1] } f^i_t \big(x \big),
\end{equation*}
where  $f^i_t \colon \mathbb{R} \to (-\infty, \infty]$ is defined by $f^i_t(x) =\infty$, $x \notin [0,1]$, and 
\begin{equation*}
	f^i_t(x)=	\mathrm{k}^i x \mathbb{E}[e^i_t] +\tfrac{\gamma^i}{2} x^2 \mathbb{E}[(e^i_t)^2] - x \lambda  \mathbb{E} \big [ e^i_t\mathds{1}_{\{ E^{\pmb{\alpha}}_T- A_T > 0\}} \big], \quad x\in [0,1].
\end{equation*}
By a classical convex analysis results, $\pmb{\alpha}^i(t)$ is a minimizer of $f^i_t$  if and only if 
\begin{equation}\label{eq:subgrad}
	\left(f^i_t \right)'\left(\pmb{\alpha}^i(t); x \right): = \lim_{\epsilon \searrow 0 }\Big( f^i_t\big(\pmb{\alpha}^i(t) +\epsilon x \big) - f^i_t\big(\pmb{\alpha}^i(t) \big)\Big)/ \epsilon \geq 0, \quad x \in \mathbb{R}.
\end{equation}
The case $x = 0$ is immediate. 
In the boundary cases $\pmb{\alpha}^i(t) = 0$ with $x < 0$ and $\pmb{\alpha}^i(t) = 1$ with $x > 0$, the derivative $\left(f^i_t\right)'(\pmb{\alpha}^i(t); x) = \infty$.
We now verify~\eqref{eq:subgrad} using~\eqref{eq:socialyoptimal} in the three remaining cases  ($i$)\;  $\pmb{\alpha}^i (t) \in (0,1)$,\; ($ii$)\; $\pmb{\alpha}^i (t) =0$ and $x>0$, and ($iii$)\;  $\pmb{\alpha}^i (t)=1$ and $x<0$. 
In each case, we have 
\begin{equation}\label{eq:subgrad2}
	\left(f^i_t \right)'(\pmb{\alpha}^i(t); x ) = 
	\mathrm{k}^i x \mathbb{E}[e^i_t] + \gamma^i \pmb{\alpha}^i (t) x \mathbb{E}[(e^i_t)^2]  - x \lambda  \mathbb{E} \big [ e^i_t\mathds{1}_{\{ E^{\pmb{\alpha}}_T- A_T > 0\}} \big].
\end{equation}
Consider the convex map $g\colon \mathbb{R}^{Tn} \to (-\infty, \infty]$ such that $g(\alpha) = R(\alpha)$, $\alpha\in \mathcal{A}$, and $g(\alpha)=\infty$, $\alpha\notin \mathcal{A}$.
Hence, by the given assumption, $\pmb{\alpha}$ is a minimizer for $g$, i.e. $g'(\pmb{\alpha}; \alpha)\geq 0$ holds for each direction $\alpha \in \mathbb{R}^{Tn}$.
Fix a direction $\alpha \in \mathbb{R}^{Tn}$ such that  $\alpha^j(s) =0$ for $s \in 0\ ..\ T-1$ and  $j\neq i$, $\alpha^i(s) =0$ for $s\neq t$, and $\alpha^i(t) = x \in \mathbb{R}$.
It then suffices to show that $  \big(f^i_t)'(\pmb{\alpha}^i(t); x ) =g'(\pmb{\alpha}; \alpha)$ holds  for each $x \in \mathbb{R}$.
The case $x = 0$ is trivial, so assume $x \ne 0$. 
Consider $\epsilon \searrow 0$ in the set $\Big(0, -\frac{\pmb{\alpha}^i(t)}{x} \vee \frac{1-\pmb{\alpha}^i(t)}{x} \Big)$, so that $\pmb{\alpha}+\epsilon \alpha \in \mathcal{A}$.
Hence, 
\begin{align}\label{eq:subgrad1}
    g'(\pmb{\alpha}; \alpha) & = \lim_{\epsilon \searrow 0} \Big( R(\pmb{\alpha}+\epsilon \alpha) -R(\pmb{\alpha})\Big) / \epsilon  = \mathrm{k}^i x \mathbb{E}[e^i_t] + \gamma^i \pmb{\alpha}^i (t) x \mathbb{E}[(e^i_t)^2] \notag  \\ 
		& + \; \lambda \lim_{\epsilon \searrow 0}\mathbb{E} \Big [ \Big( \big( E^{\pmb{\alpha}+\epsilon\alpha}_T- A_T\big)^+ -\big( E^{\pmb{\alpha}}_T- A_T\big)^+ \Big) / \epsilon \Big].
\end{align}
Let $(\epsilon_k)_{k\geq 1}$  be a sequence in $\Big(0, -\frac{\pmb{\alpha}^i(t)}{x} \vee \frac{1-\pmb{\alpha}^i(t)}{x} \Big)$ such that $\epsilon_k \searrow 0$ as $k\nearrow \infty$.
Consider 
\begin{equation*}
	X_k = \Big( \big( E^{\pmb{\alpha}+\epsilon_k\alpha}_T- A_T\big)^+ -\big( E^{\pmb{\alpha}}_T- A_T\big)^+ \Big) / \epsilon_k, \quad k\geq 1. 
\end{equation*}
Hence,  $\norm{X_k} \leq \norm{-x e^i_t}, \; k \geq 1$.
In view of assumption~\ref{ass:uniqueness}, for almost all $\omega$, 
\begin{align*}
	\lim_{k\nearrow \infty} X_k(\omega) & =  \lim_{\epsilon_k \searrow 0} \Big[ \big( E^{\pmb{\alpha}}_T(\omega)- A_T(\omega)-x\epsilon_k e^i_t(\omega) \big)^+ -\big( E^{\pmb{\alpha}}_T(\omega)- A_T(\omega)\big)^+ \Big] /\epsilon_k.\\
	& = -x e^i_t(\omega)\mathds{1}_{\big\{ E^{\pmb{\alpha}}_T- A_T > 0\big\}}(\omega).
\end{align*}
By dominated convergence theorem, we obtain 
\begin{align*}
	\lim_{\epsilon_k  \searrow 0}\mathbb{E}\left[ \Big( \big( E^{\pmb{\alpha}+\epsilon_k \alpha}_T- A_T\big)^+ -\big( E^{\pmb{\alpha}}_T- A_T\big)^+ \Big)/\epsilon_k \right] &= \lim_{k \nearrow \infty} \mathbb{E}[X_k] \\
	&=  -x \mathbb{E} \big [e^i_t\mathds{1}_{\{ E^{\pmb{\alpha}}_T- A_T  > 0\}}\big].
\end{align*}
This together with~\eqref{eq:subgrad1} implies $g'(\pmb{\alpha}; \alpha)  = \left(f^i_t \right)'(\pmb{\alpha}^i(t); x)$, and hence  $\pmb{\alpha}$ is an optimal abatement plan.

\noindent
Finally, let show that the map  $\mathbb{R}^{ n T} \ni \alpha \stackrel{R}{\mapsto} R(\alpha)$ has a unique minimizer $\pmb{\alpha} \in \mathcal{A}$.
The set $\mathcal{A}$ on which $R$ is to be minimized is non-empty, bounded and convex.	
The map $R$ is convex and lower semi-continuous.
By Theorem~27.3 in \citet[pp. 267]{rockafellar1970}, the optimization problem~\eqref{eq:socialyoptimal} admits a solution.
$R$ is strictly convex, hence the minimizer is unique.
Therefore, there exist a unique optimal abatement plan and equilibrium price. \ \finproof



\section{Explicit Formulas in a Gaussian Setup}\label{sec:senstivity}
	
In this section, we provide semi-explicit solutions in a  Gaussian set up. 
We denote by $\Phi$ and $\upphi$ the cumulative probability and the probability density  functions of the standard normal distribution and by $\mathcal{N}_d( \mu, \Gamma)$ the $d$-variate Gaussian distribution with mean $\mu$ and covariance matrix $\Gamma$.
Let $\varepsilon_{0} = (\varepsilon^0_{0}, \varepsilon^1_0,\dots, \varepsilon^n_0)^\top=0$, and $(\varepsilon_t)_{t=1}^{T-1}$ be a sequence of independent standard Gaussian vectors $\varepsilon_t = (\varepsilon^0_t, \varepsilon^1_t, \dots, \varepsilon^{n}_t)^\top$. 
We consider the natural filtration $(\mathfrak{F}_t)_{t\in 0\,..\,T-1}$ generated by the process $(\varepsilon_t)_{t=0}^{T-1}$.
We assume that emissions fluctuate over time as a result of economic shocks of different origins. To capture this, the emissions of firm $i \in {1, \dots, n}$ under the BAU case are
\begin{equation}\label{eq:normalem}
    e^i_t =  \mu^i +\sigma^i \sqrt{1-\rho^i} \;  \varepsilon^i_{t} + \sigma^i \sqrt{\rho^i} \;\varepsilon^0_{t}, \quad t \in 0\, .. \,  T-1, 
\end{equation}
where $\mu^i$, $\sigma^i$ and $\rho^i$ are positive constants such that $0<\rho^i <1$.
Note that $\mathrm{corr} (e^i_t, \varepsilon^0_t) =\sqrt{\rho^i}$, $\mathrm{corr} (e^i_t, \varepsilon^i_t) =\sqrt{1-\rho^i}$ and $\mathrm{corr} (e^i_t, e^j_t) = \sqrt{\rho^i \rho^j} $.
	
\begin{remark}
	The Gaussian distribution assumption is adopted for tractability and to obtain explicit solutions. 
	This choice is standard in economic modeling, where Gaussian shocks are widely used to capture economic fluctuations in variables such as GDP (e.g. \cite{smets2007shocks}). 
	In our context, this implies that emissions also follow Gaussian dynamics. 
	Empirical evidence from the early stages of the ETS suggests that this approximation is acceptable for well-calibrated values of $\mu^i$ and $\sigma^i$ \citep[see][]{carmona2009, aid2023,hitzemann2018}.
\end{remark}

We assume that the total emissions cap $A_T$ over the compliance period $[0, T]$ is given by
\begin{equation}\label{eq:normalcaps}
	A_T = a \left( \sum_i \sum_{t=0}^{T-1} e^i_t\right) + (1 - a)\varepsilon,
\end{equation}
where $a \in (0,1)$ is a constant, and $\varepsilon \sim \mathcal{N}_1(0,1)$ is independent of each firm's emissions processes $e^i_t$. 
The information available at time $T$ is denoted by  $\mathfrak{F}_T = \mathfrak{F}_{T-1} \vee \mathfrak{F}^\varepsilon$, which is the sigma-algebra generated by combining the information accumulated up to time $T-1$, $\mathfrak{F}_{T-1}$, with the additional information revealed by the sigma-algebra $\mathfrak{F}^\varepsilon$ generated by the random variable $\varepsilon$.
The additive term $(1 - a)\varepsilon$ is a technical adjustment introduced to satisfy Assumption~\ref{ass:uniqueness} and to guarantee the differentiability of the expected aggregated cost function $R(\alpha)$ for all values of $\alpha$, including the case $\alpha^i(t) = 1 - a$ for each $i$ and $t=0\,..\,T-1$, as required in Proposition~\ref{prop:distofcomplience}.
Within this framework, the regulator may distribute allowances proportionally to each firm's emissions under the BAU case. 
Specifically we assume,
\begin{equation*}
	A^i_t = a \sum_{s=0}^t e^i_s, \quad t \in 0\, .. \,  T-1 \quad \text{and} \quad A^i_T = a \sum_{s=0}^{T-1} e^i_s + \frac{(1 - a)\varepsilon}{n}.
\end{equation*}

\begin{remark}
    One could also accommodate time-dependent emissions parameters $\mu^i(t)$, $\sigma^i(t)$ and $\rho^i(t)$, as well as  allowance allocation parameters $a^i(t)$.
\end{remark}

\smallskip
\begin{notation}\label{not:compact}
    We define row vectors $b(t, \alpha) \in \mathbb{R}^{n+1}$ as follows: 
    \begin{align*}
         b(t,\alpha) \colon =  \Big( & \sum_{i } \sigma^i \sqrt{\rho^i}\big[1-\alpha^i (t) -a\big], \; \sigma^1  \sqrt{1-\rho^1}\big[1-\alpha^1 (t)-a\big],\\
         &  \quad \; \dots \;,   \quad \sigma^n \sqrt{1-\rho^n} \big[1-\alpha^n (t)-a \big]   \Big), \quad  t\in 1\,..\, T-1 
    \end{align*}
	and  $b(T, \alpha) := (a - 1, 0, \dots, 0) \in \mathbb{R}^{n+1}$.
	For each $t \in 1\,..\, T$, we denote by $\mathrm{B}(t, \alpha)$ the row vector (respectively, by $\mathcal{E}_t$ the column vector) obtained by concatenating the vectors $b(1, \alpha), \dots, b(t, \alpha)$ (respectively, $\varepsilon_1, \dots, \varepsilon_t$), where $\varepsilon_T \coloneqq (\varepsilon, 0, \dots, 0)^\top \in \mathbb{R}^{n+1}$.
	At time $t = 0$, we set $\mathrm{B}(0, \alpha) =0$ and $\mathcal{E}_0 =0$.
\end{notation}

Hence, the Euclidean norm $\norm{\mathrm{B}(t, \alpha)}$ of the vector $\mathrm{B}(t,\alpha)$ satisfies
\begin{equation*}
	\norm{\mathrm{B}(t, \alpha)}^2 = \sum_{s=1}^t \norm{b(s,\alpha)}^2,
\end{equation*}
where $\norm{b(s,\alpha)}$ denotes the Euclidean norm of the vector $b(s,\alpha)$.

Let $\mathrm{AC}(\alpha)$ denote the expected aggregate abatement cost, and $\mathrm{EE}(\alpha)$ denote the expected excess emissions under the abatement plan $\alpha$, defined by:
\begin{align*}
	\mathrm{AC}(\alpha) &= \sum_{i} \mathbb{E} \left[ \sum_{t=0}^{T-1} \mathrm{k}^i \alpha^i(t) e^i_t + \tfrac{\gamma^i}{2} \big( \alpha^i(t) e^i_t \big)^2 \right], \\
	\mathrm{EE}(\alpha) &= \mathbb{E} \left[ \left( E^\alpha_T - A_T \right)^+ \right].
\end{align*}
By~\eqref{eq:agregate_emision},~\eqref{eq:normalem} and~\eqref{eq:normalcaps}, 
\begin{equation*}
      Y^{\alpha}:=E^\alpha_T -A_T  
	 	=  \sum_{i } \sum_{t=0}^{T-1} \big(1-\alpha^i (t) -a \big)e^i_t  -(1-a) \varepsilon  
	 		= m(\alpha) + \mathrm{B}(T,\alpha)\mathcal{E}_T, 
\end{equation*}
where 
\begin{equation*}
	m(\alpha) := \sum_{i } \sum_{t=0}^{T-1} \Big(1-\alpha^i (t) -a \Big) \mu^i.
\end{equation*} 
Since $Y^{\alpha}$ is the linear combinations of independent Gaussian random variables, we have 
\begin{equation}\label{eq:y}
	Y^{\alpha} \sim \mathcal{N}_1 \big(m(\alpha),\norm{\mathrm{B}(T,\alpha)}^2 \big).
\end{equation}

\begin{proposition}\label{prop:distofcomplience}
   In the Gaussian setup, the expected aggregate cost $R(\alpha)$ of the  abatement plan $\alpha$ is 
   \begin{align*}
		R(\alpha) & = \mathrm{AC}(\alpha) + \lambda\  m(\alpha) \Phi\left(\frac{m(\alpha)}{\norm{\mathrm{B}(T, \alpha)}}\right) + \lambda\  \norm{\mathrm{B}(T, \alpha)} \upphi \left(\frac{m(\alpha)}{\norm{\mathrm{B}(T, \alpha)}}\right),
   \end{align*}
    where   $\mathrm{AC}(\alpha)$ is the expected aggregated abatement cost given by  
	\begin{align*}
		 & \mathrm{AC}(\alpha)  \\
		 & =  \sum_{i }\left[     \mathrm{k}^i\alpha^i (0) \mu^i +\tfrac{\gamma^i}{2}\big( \alpha^i (0) \mu^i \big)^2  + \sum_{t=1}^{T-1} \left[ \mathrm{k}^i \alpha^i (t) \mu^i +\tfrac{\gamma^i}{2} (\alpha^i (t))^2 \Big( (\mu^i)^2 +  (\sigma^i \big)^2 \Big) \right] \right]. 
	\end{align*}
\end{proposition}

\proof
The expected aggregate cost can be expressed as $R(\alpha) = \mathrm{AC}(\alpha) + \lambda\, \mathrm{EE}(\alpha)$. 
The formula for $\mathrm{AC}(\alpha)$ is straightforward.
For $X\sim \mathcal{N}_1 (\mu,\sigma^2)$, the expectation of its positive part is given by
\begin{equation*}
    \mathbb{E}[X^+]  = \mu   \Phi\Big(\frac{\mu}{\sigma}\Big) +  \sigma \upphi \Big(\frac{\mu}{\sigma}\Big). 
\end{equation*}
Applying this result with $\mu = m(\alpha)$ and $\sigma = \lVert \mathrm{B}(T,\alpha)\rVert$, and using~\eqref{eq:y} yields 
\begin{equation}\label{eq:g1}
       \mathrm{EE}(\alpha) = \mathbb{E}\left[Y^{\alpha} \right] =  m(\alpha) \Phi\left(\frac{m(\alpha)}{\norm{\mathrm{B}(T,\alpha)} }\right) +  \norm{\mathrm{B}(T,\alpha)} \upphi \left(\frac{m(\alpha)}{\norm{\mathrm{B}(T,\alpha)}}\right),
\end{equation}
which ends the proof of the proposition.\ \finproof 

\begin{proposition}\label{prop:repagent}
    The allowance price at time $t \in 0\,..\,T-1$ is 
    \begin{equation*}
        \mathbf{P}_t  =	\lambda \Phi\left( \frac{m(\pmb{\alpha}) + \mathrm{B} (t, \pmb{\alpha}) \mathcal{E}_t}{\sqrt{ \norm{\mathrm{B}(T, \pmb{\alpha})}^2- \norm{\mathrm{B}(t, \pmb{\alpha})}^2 }}  \right).
    \end{equation*}
\end{proposition}
	
\proof
By~\eqref{eq:y}, the probability density function for the cumulative net emissions $Y^{\pmb{\alpha}}$  is
\begin{equation*}
	f_{Y^{\pmb{\alpha}} }(y) = \frac{1}{\norm{\mathrm{B}(T, \pmb{\alpha})} } \upphi\left(\frac{y-m(\pmb{\alpha})}{\norm{\mathrm{B}(T, \pmb{\alpha})}}\right), \;  y \in \mathbb{R}.
\end{equation*} 
Since the allowance price $\mathbf{P}$ is the martingale closed by $\pmb{\xi}^{\pmb{\alpha}}$, it follows that  $\mathbf{P}_0 = \mathbb{E}[\pmb{\xi}^{\pmb{\alpha}}]= \lambda\ \mathbb{P} [Y^{\pmb{\alpha}}  >0 ] $ which provides the desired expression for $\mathbf{P}_0$.
At time $t\in 1\,..\, T-1$, we have   $(Y^{\pmb{\alpha}}, \mathcal{E}_t)^\top \sim \mathcal{N}_{t(n+1)+1} \Big(\mu^{(Y^{\pmb{\alpha}}, \mathcal{E}_t)}, \Gamma^{(Y^{\pmb{\alpha}}, \mathcal{E}_t)}\Big)$, with
\begin{equation*}
    \mu^{(Y^{\pmb{\alpha}}, \mathcal{E}_t)} = \Big(m(\pmb{\alpha}), 0,\dots,0\Big)^\top \quad \text{and} \quad \Gamma^{(Y^{\pmb{\alpha}},\mathcal{E}_t)} =
    \begin{bmatrix}
        \norm{\mathrm{B}(T, \pmb{\alpha})}^2 & \mathrm{B} (t, \pmb{\alpha})\\
        \mathrm{B}(t, \pmb{\alpha})^\top & I _{t(n+1)}
    \end{bmatrix},
\end{equation*}
where $I _{t(n+1)}$ is the $t(n+1)$-dimensional identity matrix.
Using standard block matrix inversion results, we obtain
\begin{multline*}
    \det\big(\Gamma^{(Y^{\pmb{\alpha}}, \mathcal{E}_t)} \big)
    =\norm{\mathrm{B}(T, \pmb{\alpha})}^2 - \norm{\mathrm{B} (t, \pmb{\alpha})}^2 \; \text{ and } 
    \\
    \Big(\Gamma^{(Y^{\pmb{\alpha}}, \mathcal{E}_t)}\Big)^{-1}= \frac{1}{\det\big(\Gamma^{(Y^{\pmb{\alpha}}, \mathcal{E}_t)} \big)}
    \begin{bmatrix}
        1 & -\mathrm{B}(t, \pmb{\alpha})\\
        -\mathrm{B} (t, \pmb{\alpha})^\top & \mathrm{B}(t, \pmb{\alpha})^\top \mathrm{B}(t, \pmb{\alpha}) + \det(\Gamma^{(Y^{\pmb{\alpha}},\mathcal{E}_t)} )I _{t(n+1)}  
    \end{bmatrix}, \hspace{1.3cm}
\end{multline*}
where $\det\big(\Gamma^{(Y^{\pmb{\alpha}}, \mathcal{E}_t)} \big)$ is the determinant of $\Gamma^{(Y^{\pmb{\alpha}}, \mathcal{E}_t)}$.	
Hence, the joint probability density function  $f_{(Y^{\pmb{\alpha}},\mathcal{E}_t)}$ of   $(Y^{\pmb{\alpha}}, \mathcal{E}_t)$ valued at $x=(y,z) \in \mathbb{R}\times \mathbb{R}^{t(n+1)}$ is 
\begin{align*}
   &  f_{(Y^{\pmb{\alpha}},\mathcal{E}_t)} (x) 
    = \frac{1}{(2\pi)^{\frac{t(n+1)+1}{2}} \sqrt{\det\big(\Gamma^{(Y^{\pmb{\alpha}},\mathcal{E}_t)} \big)}} \; \times
    \\
    & \hspace{2cm} \exp\Big(\frac{-1}{2}\big(x-\mu^{(Y^{\pmb{\alpha}},\mathcal{E}_t)}\big)^\top \big( \Gamma^{(Y^{\pmb{\alpha}},\mathcal{E}_t)}\big)^{-1}\big(x-\mu^{(Y^{\pmb{\alpha}},\mathcal{E}_t)}\big) \Big)\\
    &  = \frac{1}{(2\pi)^{\frac{t(n+1)+1}{2}} \sqrt{\det\big(\Gamma^{(Y^{\pmb{\alpha}},\mathcal{E}_t)} \big)}} \times \\
    & \exp\left( \frac{\big(y- m(\pmb{\alpha} )\big)^2 -2\mathrm{B}(t, \pmb{\alpha})z \big(y- m(\pmb{\alpha}) \big) + \big(\mathrm{B}(t, \pmb{\alpha})z\big)^2 +\det\big(\Gamma^{(Y^{\pmb{\alpha}},\mathcal{E}_t)}\big)\; \norm{z}^2 }{-2\det(\Gamma^{(Y^{\pmb{\alpha}},\mathcal{E}_t)} )} \right).
\end{align*}
Note that $\mathcal{E}_t \sim \mathcal{N}_{t(n+1) }\big(0, \emph{I}_{t(n+1)} \big)$, with density 
\begin{equation*}
    f_{\mathcal{E}_t}(z) =
    \frac{1}{(2\pi)^{\frac{t(n+1)}{2}} } \exp\left( \frac{-\norm{z}^2}{2}\right).
\end{equation*}
Hence, the conditional probability density function of $Y^{\pmb{\alpha}}$ given $\mathcal{E}_t$ is 
\begin{align*}
    & f_{Y^{\pmb{\alpha}}|\mathcal{E}_t=z}(y) 
    = f_{(Y^{\pmb{\alpha}}, \mathcal{E}_t)}(y,z)/  f_{\mathcal{E}_t}(z)\\
    & \qquad=\frac{1}{\sqrt{2\pi} \sqrt{\norm{\mathrm{B} (T, \pmb{\alpha})}^2 -\norm{\mathrm{B}(t, \pmb{\alpha})}^2} }\exp\left( -\frac{\big(y- m(\pmb{\alpha}) -\mathrm{B}(t, \pmb{\alpha})z \big)^2 }{2\norm{\mathrm{B}(T, \pmb{\alpha})}^2 -2\norm{\mathrm{B} (t, \pmb{\alpha})}^2} \right).
\end{align*}
Therefore, 
\begin{align*}
   &\mathbf{P}_t= \lambda\; \mathbb{P}_t[Y^{\pmb{\alpha}}>0] =\lambda\; \mathbb{P}[Y^{\pmb{\alpha}}>0|\mathcal{E}_t]
    =\lambda\int_{0}^\infty f_{Y^{\pmb{\alpha}}|\mathcal{E}_t}(y)dy  \\
    &  =\frac{\lambda }{\sqrt{2\pi( \norm{\mathrm{B} (T, \pmb{\alpha})}^2  -\norm{\mathrm{B} (t, \pmb{\alpha})}^2) }  } \int_{0}^\infty \exp\Bigg( \frac{-\left(y- m(\pmb{\alpha}) -\mathrm{B}(t, \pmb{\alpha})\mathcal{E}_t\right)^2 }{2(\norm{\mathrm{B} (T, \pmb{\alpha})}^2 -\norm{\mathrm{B} (t, \pmb{\alpha})}^2)} \Bigg)dy.
\end{align*}
A change of variable $w =\left(y- m(\pmb{\alpha}) -\mathrm{B}(t, \pmb{\alpha})\mathcal{E}_t\right)\big/\sqrt{\norm{\mathrm{B} (T, \pmb{\alpha})}^2 -\norm{\mathrm{B} (t, \pmb{\alpha})}^2}$, yields 
\begin{align*}
   &\mathbf{P}_t  = \lambda\frac{1 }{\sqrt{2\pi} } \int_{w(0)}^\infty \exp\left( \frac{-w^2 }{2} \right)dw = \lambda \left[ 1- \Phi\left( w(0) \right) \right] = \lambda \Phi\left( -w(0) \right),
\end{align*}
which is  the desired expression of $\mathbf{P}_t$.\ \finproof\\

Proposition~\ref{prop:repagent} highlights two central features of the allowance price $\mathbf{P}_t$ at times $t\in 0\,..\, T-1$. 
First, the term $m(\pmb{\alpha})$ represents the expected value of cumulative net emissions for the abatement plan $\pmb{\alpha}$, while the product $\mathrm{B}(t, \pmb{\alpha})\mathcal{E}_t = \sum_{s=0}^t b(s,\pmb{\alpha}) \varepsilon_s$ captures the cumulative effect of economic shocks observed up to time $t$, weighted by firms' exposure $b(s,\pmb{\alpha})$ to risk and the extent to which their abatement efforts differ from the regulatory cap $a$. 
Second, the expression $\sqrt{\norm{\mathrm{B}(T, \pmb{\alpha})}^2 - \norm{\mathrm{B}(t, \pmb{\alpha})}^2}=\sqrt{\sum_{s=t+1}^{T-1} \norm{b(s,\pmb{\alpha})}^2} $ measures the residual uncertainty about future economic activity.
A higher value of this term reflects greater uncertainty about future economic activities, which typically lowers the current price of emission allowances. 
Conversely, when future economic conditions are more predictable, the residual uncertainty is smaller, leading to higher allowance prices.

\begin{proposition}\label{prop:mean_var}
    The expected value $\mu_{\mathbf{P}_t}$ and variance $\sigma^2_{\mathbf{P}_t} $ of the allowance price $\mathbf{P}_t$ at time $t \in 0\;..\; T$ are given by
    \begin{align*}
         & \mu_{\mathbf{P}_t} = \lambda \; \Phi\left(\frac{m(\pmb{\alpha})}{\norm{\mathrm{B}(T, \pmb{\alpha})}}\right) = \mathbf{P}_{0}, \\
         		 & \sigma^2_{\mathbf{P}_t} 
				 =\frac{\lambda^2}{\pi} \int_{\ell(t, \pmb{\alpha})}^1 \frac{\exp\left(-\frac{1}{2} \left(\frac{m(\pmb{\alpha})}{\norm{\mathrm{B}(T, \pmb{\alpha})}}\right)^2 (1 + u^2)\right)}{1 + u^2} \, du,
    \end{align*}
	where 
	\begin{equation*}
		\ell(t, \pmb{\alpha}) =\sqrt{ \frac{\norm{\mathrm{B}(T, \pmb{\alpha})}^2- \norm{\mathrm{B}(t, \pmb{\alpha})}^2}{ \norm{\mathrm{B}(T, \pmb{\alpha})}^2 +\norm{\mathrm{B} (t, \pmb{\alpha})}^2 }}.
	\end{equation*}
\end{proposition}
	
\proof
Given~\eqref{eq:y}, Proposition~\ref{prop:Fallocation}  together with~\eqref{eq:terminalprice}  implies 
\begin{equation*}
    \mu_{\mathbf{P}_T} = \mathbb{E}[\mathbf{P}_T] = \lambda\; \mathbb{P}\big[E^{\pmb{\alpha}}_T- A_T> 0\big]= \lambda\;\Phi\left(\frac{m(\pmb{\alpha})}{\norm{\mathrm{B}(T, \pmb{\alpha})}}\right).
\end{equation*}
By the martingale property of the price process $\mathbf{P}$, it follows that $\mu_{\mathbf{P}_t} = \mathbf{P}_0 = \mu_{\mathbf{P}_T}$ for $t \in 0\;..\; T$.
Let $X \sim \mathcal{N}_1(\mu, \sigma^2)$, and let $Z_1$ and $Z_2$ denote two independent standard normal random variables, each independent of $X$. Define $Y_i = Z_i - X$ for $i=1,2$.
Then, 
\begin{align}
    \mathbb{E}[\Phi^2(X)] & = \mathbb{P} [Z_1 \leq X, Z_2 \leq X ]  = \mathbb{P}[Y_1\leq 0, Y_2 \leq 0 ] \notag\\
		& = \Phi_2\left( \frac{\mu}{\sqrt{1+\sigma^2}}, \frac{\mu}{\sqrt{1+\sigma^2} }; \frac{\sigma^2}{1+\sigma^2} \right) \notag\\
			& =  \Phi\left(  \frac{\mu}{\sqrt{1+\sigma^2}}\right) - 2\mathcal{T}\left(\frac{\mu}{\sqrt{1+\sigma^2} }, \frac{1}{\sqrt{1+2\sigma^2}} \right),\label{eq:ownt}
\end{align}
where $\Phi_2(x,y; \rho)$ is the cdf of bi-variate standard normal distribution with correlation $\rho$, $\mathcal{T}$ is the Owen's $T$-function given by 
\begin{equation*}
		\mathcal{T}(h, d) = \frac{1}{2\pi} \int_0^d \frac{\exp\left(-\frac{1}{2} h^2(1 + u^2)\right)}{1 + u^2} du,
\end{equation*}
and the last equality holds due to \citep[Eqn. $3.5$]{Owen1980}. 
For $ t\in 1\,..\, T-1$, in Proposition~\ref{prop:repagent}, the random variable 
\begin{multline*}
    X_t \colon=  \frac{m(\pmb{\alpha}) + \mathrm{B} (t, \pmb{\alpha}) \mathcal{E}_t}{\sqrt{ \norm{\mathrm{B}(T, \pmb{\alpha})}^2- \norm{\mathrm{B}(t, \pmb{\alpha} )}^2}} \\
    \sim 
    \mathcal{N}_1 \left( \frac{m(\pmb{\alpha}) }{\sqrt{\norm{\mathrm{B}(T, \pmb{\alpha})}^2- \norm{\mathrm{B}(t, \pmb{\alpha})}^2}}, \; \frac{\norm{\mathrm{B} (t, \pmb{\alpha})}^2}{\norm{\mathrm{B}(T, \pmb{\alpha})}^2- \norm{\mathrm{B}(t, \pmb{\alpha})}^2}\right).
\end{multline*}
This together with Eqn.~\eqref{eq:ownt} implies
\begin{equation*}
	\mathbb{E}[\Phi^2(X_t)] = \Phi\left( \frac{m(\pmb{\alpha})}{\norm{\mathrm{B}(T, \pmb{\alpha})}}\right) - 2\mathcal{T}\left( \frac{m(\pmb{\alpha})}{\norm{\mathrm{B}(T, \pmb{\alpha})}}, \ell(t, \pmb{\alpha})  \; \right).
\end{equation*}
An application of the identity $\Phi^2(h) = \Phi(h) -2\mathcal{T}(h,1)$ ( see \citet[Eqn.~2.3]{Owen1980} ) yields 
\begin{equation*}
	\mathbf{P}^2_0 =\lambda^2 \Phi^2\left(\frac{m(\pmb{\alpha})}{\norm{\mathrm{B} (T, \pmb{\alpha})}}\right)  = \lambda^2 \Phi\left(\frac{m(\pmb{\alpha})}{\norm{\mathrm{B} (T, \pmb{\alpha})}}\right) -2\lambda^2\mathcal{T}\left(\frac{m(\pmb{\alpha})}{\norm{\mathrm{B} (T, \pmb{\alpha})}}, 1\right). 
\end{equation*}
In view of Proposition~\ref{prop:repagent}, it holds 
\begin{equation*}
   \sigma^2_{\mathbf{P}_t} = \mathbb{E}[\mathbf{P}^2_t] -\mu^2_{\mathbf{P}_t}
    = \lambda^2 \;	\mathbb{E}[\Phi^2(X_t)]-\mathbf{P}^2_0,
\end{equation*}
which gives the required expression of $ \sigma^2_{\mathbf{P}_t}$, $ t\in 1\,..\, T-1$. 
For $t = 0$, the allowance price $\mathbf{P}_0$ is deterministic, implying zero variance. 
As $\ell(0,\pmb{\alpha})=1$, the result follows directly for $t=0$.
For $t=T$, we obtain  
\begin{align*}
     \sigma^2_{\mathbf{P}_T} & = \mathbb{E}[\mathbf{P}^2_T]-\mu^2_{\mathbf{P}_T} = \lambda^2\; \mathbb{P}\big[E^{\pmb{\alpha}}_T-A_T > 0\big] -\mathbf{P}^2_0 \\
   	 & = \lambda^2 \; \Phi\left(-\frac{m(\pmb{\alpha})}{\norm{\mathrm{B} (T, \pmb{\alpha})}}\right)  \Phi\left(\frac{m(\pmb{\alpha})}{\norm{\mathrm{B} (T, \pmb{\alpha})}}\right) \\
	 & = \lambda^2 \left[ \Phi\left(\frac{m(\pmb{\alpha})}{\norm{\mathrm{B} (T, \pmb{\alpha})}}\right) -\Phi^2\left(\frac{m(\pmb{\alpha})}{\norm{\mathrm{B} (T, \pmb{\alpha})}}\right)\right].  
\end{align*} 
Using the identity $\Phi^2(h) = \Phi(h) - 2\mathcal{T}(h, 1)$, we obtain the desired expression for $\sigma^2_{\mathbf{P}_T}$.\ \finproof\\

The above results characterize the dynamics of the allowance price $\mathbf{P}_t$ over time. 
Both its expected value and variance are governed by the signal-to-noise ratio $m(\pmb{\alpha})/\norm{\mathrm{B}(T, \pmb{\alpha})}$ of the cumulative net emissions, which quantifies the uncertainty surrounding regulatory compliance at the terminal date.
A higher signal-to-noise ratio indicates that the likelihood of exceeding the emissions cap $a$ is more clearly identifiable, resulting in a higher expected allowance price and lower price volatility. 
A tighter emissions cap  (lower $a$) similarly heightens the clarity of compliance risk, producing comparable effects. 
Conversely, a lower signal-to-noise ratio reflects increased uncertainty stemming from stochastic emissions, thereby leading to greater price volatility and a reduced expected price. 
The lower limit $\ell(t, \pmb{\alpha})$ of the integral in the expression for the variance $\sigma^2_{\mathbf{P}_t}$ reflects the evolving trade-off between residual uncertainty about future economic activity measured by $\norm{\mathrm{B}(T, \pmb{\alpha})}^2 - \norm{\mathrm{B}(t, \pmb{\alpha})}^2$ and the information accumulated up to time $t$, captured by $\norm{\mathrm{B}(t, \pmb{\alpha})}^2$.
As time advances, the lower bound of the integral decreases, leading to an increase in price volatility. 
At the terminal time $T$, this bound converges to zero, reflecting the resolution of all uncertainty.

\section{Sensitivity Analysis}\label{sec:theoretical_sensitivity}

Although the effects of key parameters, such as the penalty rate $\lambda$, the emissions cap $a$, and the characteristics of baseline emissions $\mu^i$, $\sigma^i$, and $\rho^i$, can be partially understood from the characterization of the allowance price $\mathbf{P}_t$ in Proposition~\ref{prop:repagent}, the impact of the abatement cost parameters is less transparent, as they do not explicitly appear in the analytical expression for $\mathbf{P}_t$.
Consequently, the precise manner in which the full set of model parameters influences the allowance price is only partially understood. 
To analyze these interdependencies, we undertake a sensitivity analysis based on the implicit function theorem.

For analytical tractability and to avoid the complications associated with Lagrange multipliers, we henceforth assume that the penalty level $\lambda$ is chosen such that the optimal abatement plan lies strictly within the interior of the admissible set, i.e. $\pmb{\alpha} \in \mathrm{int}(\mathcal{A})$.
Under this assumption, and since the model parameters $\mathrm{k}^i$,  $\gamma^i$, $\mu^i$, $\sigma^i$,  $\rho^i$ and $a$ are time-invariant, the optimal abatement plans are constant across all periods $t \in 1\,..\,T-1$, with the possible exception at $t = 0$, as formalized in the following proposition.

\begin{proposition}\label{prop:equal_efoort}
	For each firm $i$, the optimal abatement plans remain constant at times  $t \in  1\,..\,T-1$, i.e.\ 
	\begin{equation*}
		\pmb{\alpha}^i(t)=\pmb{\alpha}^i(1), \quad t \in 1\,..\,T-1.
	\end{equation*}
\end{proposition}

\proof See Section~\ref{app:equal_efoort}. 

In light of the preceding result, the optimal abatement plan $\pmb{\alpha}$ can be represented as
\begin{equation*}
	\pmb{\alpha} = \displaystyle\left(\pmb{\alpha}^1(0), \pmb{\alpha}^1(1),\dots, \pmb{\alpha}^n(0), \pmb{\alpha}^n(1) \right)^\top \in \mathcal{A} \subset \mathbb{R}^{2n},
\end{equation*}
with the associated aggregated cost function
\begin{equation}\label{eq:agregate_cost}
	R(\alpha) = \mathrm{AC}(\alpha ) + \lambda\; m(\alpha) \Phi\left(\frac{m(\alpha)}{\norm{\mathrm{B}(T,\alpha)} }\right) +  \lambda\; \norm{\mathrm{B}(T,\alpha)} \upphi \left(\frac{m(\alpha)}{\norm{\mathrm{B}(T,\alpha)}}\right)
\end{equation} 
defined on $\mathcal{A} \subset \mathbb{R}^{2n}$, where 
\begin{align*}
	& \mathrm{AC}(\alpha)  =  \sum_{i } \left[ \mathrm{k}^i\alpha^i (0) \mu^i +\tfrac{\gamma^i}{2}( \alpha^i (0) \mu^i)^2  \right]\\ 
		& \qquad \qquad  +\;  (T-1) \sum_i  \left[ \mathrm{k}^i \alpha^i (1) \mu^i +\tfrac{\gamma^i}{2} (\alpha^i (1))^2 \big[ (\mu^i)^2 +  (\sigma^i)^2 \big] \right], \\
			& m(\alpha) = \sum_{i } \Big[T(1-a) -\alpha^i(0) -(T-1)\alpha^i(1) \Big] \mu^i \quad \text{and} \\
        	& \norm{\mathrm{B}(T,\alpha)}^2  =  (T-1)\norm{b(1,\alpha)}^2 + (1-a)^2. 
\end{align*}
Consequently, the results of Propositions~\ref{prop:repagent} and \ref{prop:mean_var} reduce to the following simplified forms:
\begin{corollary}
	The allowance price $\mathbf{P}_t$ and its variance $\sigma^2_{\mathbf{P}_t}$ simplified to 
	\begin{align*}
	& \mathbf{P}_t   =	\lambda \Phi\left( \dfrac{m(\pmb{\alpha}) +  b(1,\pmb{\alpha}) \sum_{s=0}^t \varepsilon_s }{\sqrt{ (T-1-t) \norm{ b(1, \pmb{\alpha}) }^2 +(1-a)^2 }}  \right),  \quad t\in 0\,..\,T-1, \\
		 & \sigma^2_{\mathbf{P}_t}=\frac{\lambda^2}{\pi} \int_{\ell(t, \pmb{\alpha})}^1 \frac{\exp\left(-\frac{1}{2} \left(\frac{m(\pmb{\alpha})}{{\sqrt{ (T-1) \norm{ b(1, \pmb{\alpha}) }^2 +(1-a)^2 }}}\right)^2 (1 + u^2)\right)}{1 + u^2} \, du, \;  t \in 0\;..\; T, 
\end{align*}
where 
\begin{equation*}
	\ell(t, \pmb{\alpha})  =
	\begin{cases}
		\sqrt{ \frac{(T-1-t)\norm{b(1,\pmb{\alpha})}^2  + (1-a)^2 }{(T-1+t)\norm{b(1,\pmb{\alpha})}^2  + (1-a)^2}},  &  t \in 0\;..\; T-1\\
		0, & t=T. 
	\end{cases}
\end{equation*}
\end{corollary}

\proof 
By Proposition~\ref{prop:equal_efoort}, the optimal abatement plans are constant across periods $t \in 1\;..\; T-1$, which implies $b(t,\pmb{\alpha}) = b(1, \pmb{\alpha} )$, $t \in 1\;..\; T-1$, $\norm{\mathrm{B}(T, \pmb{\alpha})}^2 = (T-1)\norm{b(1,\pmb{\alpha})}^2 + (1 - a)^2$ and $\norm{\mathrm{B}(t, \pmb{\alpha})}^2 = (T - t)\norm{b(1,\pmb{\alpha})}^2$. 
Substituting these expressions into the formulas of allowance price $\mathbf{P}_t$  and its variance  $\sigma_{\mathbf{P}_t}$ from Propositions~\ref{prop:repagent} and \ref{prop:mean_var} yields the claimed result directly.\ \finproof

\begin{notation}
	We denote by $\partial_x f$ the Jacobian matrix of a vector-valued function $f$, and by $\nabla_x f$ and $\nabla_x^2 f$ the gradient vector and Hessian matrix, respectively, of a real-valued function $f$.
\end{notation}

The first-order condition associated with the minimization problem~\eqref{eq:socialyoptimal} is
\begin{equation}\label{eq:foc}
	 \nabla_\alpha R(\pmb{\alpha})=0.
\end{equation}
In view of  \eqref{eq:agregate_cost} and Lemma \ref{lem:lemakkt} in Section~\ref{app:equal_efoort}, the first order condition~\eqref{eq:foc} simplifies to: 
\begin{align*}
	& \mathrm{k}^i \mu^i + \gamma^i (\mu^i)^2  \pmb{\alpha}^i (0)-  \lambda \;\Phi\left(\frac{m(\pmb{\alpha})}{\norm{\mathrm{B}(T,\alpha)}}\right) \mu^i = 0,  \quad i \in 1\,..\,n,\\
		& \mathrm{k}^i \mu^i + \gamma^i\big[ (\mu^i)^2+(\sigma^i)^2 \big] \pmb{\alpha}^i (1)- \lambda \;\Phi\left(\frac{m(\pmb{\alpha})}{\norm{\mathrm{B}(T,\alpha)}}\right)  \mu^i \\
		& \hspace{4cm} -\; \frac{\lambda }{\norm{\mathrm{B}(T, \pmb{\alpha})} } \upphi\left(\frac{m(\pmb{\alpha})}{\norm{\mathrm{B}(T,\pmb{\alpha})}}\right) r^i(\pmb{\alpha})= 0,  \quad i \in 1\,..\,n,
\end{align*}
where 
	\begin{equation*}
     	r^i(\pmb{\alpha}) = (\sigma^i)^2 (1-\rho^i) \big[1-\pmb{\alpha}^i (1)-a \big] + \sigma^i \sqrt{\rho^i} \; \sum_{j} \sigma^j \sqrt{\rho^j}\;\big[1-\pmb{\alpha}^j(1)-a \big].
	\end{equation*}

Suppose  $\theta = (\lambda, (\mathrm{k}^i)_{i=1}^n, (\gamma^i)_i, (\mu^i)_i, (\sigma^i)_i, (\rho^i)_i,a)^\top$ be a vector of exogenous parameters.
To emphasize the dependence of an arbitrary  function $f(\alpha)$ on these parameters, we henceforth write $f(\alpha; \theta)$. 
Consider 
\begin{equation}\label{eq:funjac_n}
	\mathrm{int}(\mathcal{A})  \times \mathcal{V} \ni (\alpha; \theta) \stackrel{f}{\mapsto}  \nabla_\alpha R(\alpha; \theta) \in \mathbb{R}^{2n}, 
\end{equation}
where $\mathcal{V} = (0,\infty)^{4n+1}\times  (0, 1)^{n+1}$.
The Jacobian matrices of $f$ with respect to $\alpha$ and  $\theta$ are given by 
\begin{equation*}
	\nabla^2_\alpha R (\alpha; \theta)  \qquad
	 \text{and} \qquad \partial_{\theta} \nabla_\alpha R(\alpha ;\theta).
\end{equation*}

\begin{proposition}\label{prop:implicitfun_n}
	Let $\theta \in \mathcal{V}$ denote a fixed vector of exogenous parameters, and let $\pmb{\alpha} \in \mathrm{int}(\mathcal{A})$ be the corresponding optimal abatement plan. 
	Suppose that the Hessian $\nabla^2_\alpha R(\pmb{\alpha}; \theta)$ is invertible. 
	Then, for any neighborhood $\mathcal{O}_{\pmb{\alpha}} \subseteq \mathrm{int}(\mathcal{A})$ of $\pmb{\alpha}$, there exists a neighborhood $\mathcal{O}_{\theta} \subseteq \mathcal{V}$ of $\theta$ and a continuously differentiable mapping $\pmb{\alpha} (\cdot): \mathcal{O}_{\theta} \to \mathcal{O}_{\pmb{\alpha}}$ such that $\pmb{\alpha}(\vartheta)$ is an optimal abatement plan for every $\vartheta \in \mathcal{O}_{\theta}$.

	\medskip
	\noindent
	Moreover, the following hold:

	\begin{itemize}[fullwidth]
	\item[\textbf{(i)}] The sensitivity of the optimal abatement plan $\pmb{\alpha}$ with respect to the parameters satisfies
	\begin{equation*}
    	\partial_{\theta}\pmb{\alpha}(\theta)
    	= - \big( \nabla^2_\alpha R(\pmb{\alpha}(\theta); \theta)\big)^{-1} 
      \, \partial_{\theta} \nabla_\alpha R(\pmb{\alpha}(\theta); \theta).
	\end{equation*}

	\item[\textbf{(ii)}] The price $\mathbf{P}_t\big(\pmb{\alpha}(\theta)\big)$ at time $t \in 0\,..\,T-1$  is differentiable in $\theta$ and 
	\begin{equation*}
    	\partial_{\theta} \mathbf{P}_t \big(\pmb{\alpha}(\theta)\big) 
    	= \partial_{\theta} g(t, \pmb{\alpha}(\theta); \theta) 
    	+ \partial_\alpha g(t, \pmb{\alpha}(\theta); \theta)\; \partial_{\theta}\pmb{\alpha}(\theta),
	\end{equation*}
	where
	\begin{equation*}
    	g(t, \alpha; \theta) = \lambda \Phi\left( 
        	\frac{ m(\alpha; \theta) +  b(1, \alpha; \theta) \sum_{s=0}^t \varepsilon_s }
             { \sqrt{ (T - 1 - t)\, \norm{b(1, \alpha; \theta) }^2 + (1 - a)^2 } }
    	\right).
	\end{equation*}

	\item[\textbf{(iii)}] The variance $\sigma^2_{\mathbf{P}_t}(\pmb{\alpha}(\theta))$ of the price at time $t \in 0\,..\,T$ is differentiable in $\theta$ and 
	\begin{multline*}
	\partial_{\theta} \sigma^2_{\mathbf{P}_t}\big(\pmb{\alpha}(\theta)\big) 
	=
	\int_{\ell(t, \pmb{\alpha}(\theta); \theta)}^1 
	\left[
    	\partial_{\theta} h(u, \pmb{\alpha}(\theta); \theta) 
    	+ \partial_\alpha h(u, \pmb{\alpha}(\theta); \theta)\; \partial_{\theta}\pmb{\alpha}(\theta)
	\right] du \\
	-\, h\big(\ell(t, \pmb{\alpha}(\theta); \theta), \pmb{\alpha}(\theta); \theta\big) 
	\left[
    	\partial_{\theta} \ell(t, \pmb{\alpha}(\theta); \theta) 
    	+ \partial_\alpha \ell(t, \pmb{\alpha}(\theta); \theta)\; \partial_{\theta}\pmb{\alpha}(\theta) 
	\right],
	\end{multline*}
	where
	\begin{equation*}
    	h(u, \alpha; \theta) = \frac{\lambda^2}{\pi(1 + u^2)} \,
    	\exp\left(
        -\frac{1}{2} 
        	\left(
            	\frac{ m(\alpha; \theta) }{ \sqrt{ (T - 1)\, \norm{b(1, \alpha; \theta)}^2 + (1 - a)^2 } }
        	\right)^2 (1 + u^2)
    	\right).
	\end{equation*}
	\end{itemize}
\end{proposition}
	
\proof
The function $f$ in~\eqref{eq:funjac_n} is continuously differentiable on $\mathrm{int}(\mathcal{A}) \times \mathcal{V}$. The first-order optimality condition in~\eqref{eq:foc} implies that $f\big(\pmb{\alpha}; \theta \big) = 0$. By assumption, the Jacobian of $f$ with respect to $\alpha$, namely $\nabla^2_\alpha R(\pmb{\alpha}; \theta)$, is invertible.
Therefore, the function $f$ satisfies the conditions of the implicit function theorem (see e.g. \citet[Theorem 4.B]{zeidlernonlinear1}), which establishes the existence of a continuously differentiable mapping $\pmb{\alpha}(\theta)$, as well as the expression for its derivative given in part (i) of the proposition.
To establish parts (ii) and (iii), we use the formula for $\partial_{\theta} \pmb{\alpha}(\theta)$ derived in (i), together with the chain rule and the Leibniz rule for differentiation under the integral sign. 
\ \finproof\\
	
The $\theta$-elasticity of a function $f$ is defined as
\begin{equation}\label{eq:elast}
	\eta_\theta f = \frac{\partial f}{\partial \theta} \Big/ \left( \frac{f}{\theta} \right),
\end{equation}
which quantifies the proportional change in $f$ resulting from a proportional change in the parameter $\theta$.
Proposition~\ref{prop:implicitfun_n} enables the quantification of the elasticity of abatement plans $\pmb{\alpha}$, the allowance price $\mathbf{P}_t$, and its standard deviation $\sigma_{\mathbf{P}_t}$ with respect to variations in the underlying model parameters.
Numerical illustrations of these elasticity measures are presented in Section~\ref{sec:numerical}.

\section{Numerical Case Study}\label{sec:numerical}

In this section, we analyze the properties of the model numerically. 
The time horizon $T$, interpreted as the duration of the compliance period, is set to five years. 
We adopt a monthly time resolution, dividing each year into twelve periods. This results in a simulation time path of $t \in 0\,..\, 60$ months.

\begin{table}[htp]
	\caption{Summary of the baseline input parameter values.  
	Linear abatement cost coefficients are $\mathrm{k}^i$ (€/ton) and quadratic coefficients are $\gamma^i$ (€/$10^6$ ton$^2$), shown in parentheses.}
	\label{tab:parameters}
	\centering
	\begin{tabular}{@{}llr@{}} 
		\toprule
		\textbf{Parameters}  & \textbf{Description}  & \textbf{Value} \\[0.6ex]
		\midrule
		\multicolumn{3}{@{}l}{\textbf{\emph{General model parameters}}} \\[0.6ex]
		$T$   & End of the compliance period  & $5$ years \\				
		$\lambda$  & Penalty price per ton of excess emission  & 100€\\
		$\theta$  & Emissions cap percentage set by regulator  & $0.49$ \\ 
		$\mu$ & Mean of each firm's BAU emissions  & $2.71\times 10^7$ tons \\
		$\sigma$ & Std. deviation of each firm's BAU emissions  & $7.86 \times 10^6$ tons \\
		$\rho$	& Correlation between firms' emissions	& $0.85$ \\[0.6ex]

		\multicolumn{3}{@{}l}{\textbf{\emph{Firm-specific abatement cost parameters}}} \\[0.6ex]
		$\mathrm{k}^{c_1}$ ($\gamma^{c_1}$) & Linear (quadratic) cost coefficient for firm $c_1$       & 20 (2.38) \\
		$\mathrm{k}^{c_2}$ ($\gamma^{c_2}$) & Linear (quadratic) cost coefficient for firm $c_2$       & 25 (2.43) \\
		$\mathrm{k}^{c_3}$ ($\gamma^{c_3}$) & Linear (quadratic) cost coefficient for firm $c_3$       & 30 (2.48) \\
		$\mathrm{k}^{c_4}$ ($\gamma^{c_4}$) & Linear (quadratic) cost coefficient for firm $c_4$       & 35 (2.53) \\
		$\mathrm{k}^{d_1}$ ($\gamma^{d_1}$) & Linear (quadratic) cost coefficient for firm $d_1$       & 40 (2.58) \\
		$\mathrm{k}^{d_2}$ ($\gamma^{d_2}$) & Linear (quadratic) cost coefficient for firm $d_2$       & 45 (2.63) \\
		$\mathrm{k}^{d_3}$ ($\gamma^{d_3}$) & Linear (quadratic) cost coefficient for firm $d_3$       & 50 (2.68) \\
		$\mathrm{k}^{d_4}$ ($\gamma^{d_4}$) & Linear (quadratic) cost coefficient for firm $d_4$       & 55 (2.73) \\
		\bottomrule              
	\end{tabular}
\end{table}

Throughout the analysis, we consider eight groups of firms that share identical emission parameters, i.e. $\mu^i = \mu$, $\sigma^i = \sigma$ and $\rho^i = \rho$ for all $i$.
While emissions parameters are uniform across firms, abatement cost parameters remain heterogeneous, reflecting firm-specific differences in abatement technologies.
Based on their cost structures, firms are grouped into eight types, ranging from the most cost-effective (\emph{cheap}, $c_1$--$c_4$) to the most costly (\emph{dear}, $d_1$--$d_4$).
The \emph{cheap} types mostly correspond to firms in the power sector, whereas the \emph{dear} types correspond to those in the industrial sector.

In view of~\eqref{eq:normalem}, the aggregate  emissions $\sum_i \sum_{t=0}^{T-1} e^i_t$ under the BAU case follow a normal distribution, $\mathcal{N}_1(\bar{\mu}, \bar{\sigma}^2)$, where  $\bar{\mu} = nT\mu$ and $\bar{\sigma}^2 = n \sigma^2 (T-1)(1 - \rho + n\rho)$, with $n = 8$ number of firms.
According to \citet[pp. 154]{mckinesy2009}, the EU's BAU emissions for 2030 are projected at $4.9 \times 10^9$ tons, with an average abatement potential of 51\%. 
The power and industry sectors together account for 53\% of these emissions (27\% and 26\%, respectively), aligning the model's scope with roughly half of the EU's expected total. 
Accordingly, we set $\bar{\mu}= 13\times 10^9$ tons and $a  = 0.49$. 
To capture emissions volatility, we adopt the calibration from \citet{aid2023}, where annual EU emissions have a standard deviation of $2\times 10^8$ tons and a correlation coefficient with a common driving factor of $0.92$. 
Thus, the standard deviation of cumulative emissions over the compliance period is set to $\bar{\sigma} = \sqrt{T}\times 2 \times 10^8 \text{ tons} = 4.5 \times 10^8 \text{ tons}$, while the correlation parameter is specified as $\sqrt{\rho} = 0.92$.
This  gives the mean of firm-level BAU emissions $\mu$ and standard deviation $\sigma$  as shown in Table~\ref{tab:parameters}. 

In line with current ETS regulations, the penalty level is set to $\lambda = 100$€  per ton of carbon. 
For the purpose of our analysis, the linear abatement cost coefficients $\mathrm{k}^i$ are fixed as specified in Table~\ref{tab:parameters}, while the quadratic coefficients $\gamma^i$ are calibrated to ensure that the initial allowance price satisfies $\mathbf{P}_0 = 75$€, consistent with the market price observed in 2024.
Table~\ref{tab:parameters} presents a summary of the baseline parameter values employed in our computations.

The remainder of this section investigates the influence of individual components of the parameter vector $\theta = (\lambda, (\mathrm{k}^i)_{i=1}^n, (\gamma^i)_{i=1}^n, \mu, \sigma, \rho, a)^\top$ on key market outcomes. 
In particular, we analyze the sensitivity of the average allowance price $\mu_{\mathbf{P}_t}$, the standard deviation of the allowance price $\sigma_{\mathbf{P}_t}$, the average abatement efforts $\pmb{\alpha}$, and the expected excess emissions $\mathrm{EE}$. 
This analysis relies on elasticity measures as defined in Eqn.~\eqref{eq:elast}, facilitating a comparison of the relative impact of each parameter.

\subsection{Dependence on the Regulatory Parameters}

The penalty rate $\lambda$ and the emissions cap $a$ are key policy instruments, whose calibration typically reflects  political decisions.
This section investigates the impact of varying these parameters on key market outcomes.

\begin{table}[htp!]
	\centering
     \caption{The expected excess emissions $\mathrm{EE}$ (in $10^6$ tons), the expected allowance prices $\mu_{\mathbf{P}_t}$ (€), and \
     the elasticities of $\mu_{\mathbf{P}_t}$ and abatement plans $\pmb{\alpha}^i(t)$ with respect to penalty $\lambda$ (€)  (left panel)  and cap $a$ (right panel).}
    	\label{tab:lambda_theta_sidebyside}
	\begin{minipage}[t]{0.42\textwidth}
		\centering
		\begin{tabular}{lrrrr}
            \multicolumn{5}{@{}c}{ \textbf{Dependence on penalty $\lambda$}  }\\
			\toprule
			$\lambda$ & $70$ & $75$ & $90$ & $100$ \\
			\midrule
$\mathrm{EE}$ & 889.54 & 35.60 & 12.20 & 9.48 \\
$\mu_{\mathbf{P}_t}$ & 70.00 & 74.91 & 75.01 & 75.02 \\
\midrule
$\eta_{\lambda}(\mu_{\mathbf{P}_t})$ & 1.00 & 0.18 & 0.00 & 0.00 \\
$\eta_{\lambda}(\pmb{\alpha}^{c_1}(0))$ & 1.40 & 0.24 & 0.00 & 0.00 \\
$\eta_{\lambda}(\pmb{\alpha}^{c_1}(1))$ & 1.40 & 0.23 & -0.01 & -0.01 \\
$\eta_{\lambda}(\pmb{\alpha}^{c_2}(0))$ & 1.56 & 0.27 & 0.00 & 0.00 \\
$\eta_{\lambda}(\pmb{\alpha}^{c_2}(1))$ & 1.56 & 0.26 & -0.01 & -0.01 \\
$\eta_{\lambda}(\pmb{\alpha}^{c_3}(0))$ & 1.75 & 0.30 & 0.00 & 0.00 \\
$\eta_{\lambda}(\pmb{\alpha}^{c_3}(1))$ & 1.75 & 0.30 & -0.00 & -0.00 \\
$\eta_{\lambda}(\pmb{\alpha}^{c_4}(0))$ & 2.00 & 0.33 & 0.00 & 0.00 \\
$\eta_{\lambda}(\pmb{\alpha}^{c_4}(1))$ & 2.00 & 0.35 & 0.00 & 0.00 \\
$\eta_{\lambda}(\pmb{\alpha}^{d_1}(0))$ & 2.33 & 0.38 & 0.00 & 0.00 \\
$\eta_{\lambda}(\pmb{\alpha}^{d_1}(1))$ & 2.33 & 0.41 & 0.01 & 0.01 \\
$\eta_{\lambda}(\pmb{\alpha}^{d_2}(0))$ & 2.80 & 0.45 & 0.00 & 0.00 \\
$\eta_{\lambda}(\pmb{\alpha}^{d_2}(1))$ & 2.80 & 0.50 & 0.02 & 0.02 \\
$\eta_{\lambda}(\pmb{\alpha}^{d_3}(0))$ & 3.50 & 0.54 & 0.01 & 0.00 \\
$\eta_{\lambda}(\pmb{\alpha}^{d_3}(1))$ & 3.50 & 0.61 & 0.03 & 0.03 \\
$\eta_{\lambda}(\pmb{\alpha}^{d_4}(0))$ & 4.67 & 0.67 & 0.01 & 0.01 \\
$\eta_{\lambda}(\pmb{\alpha}^{d_4}(1))$ & 4.67 & 0.78 & 0.05 & 0.04 \\
			\bottomrule
		\end{tabular}
	\end{minipage}
	\hfill
	\begin{minipage}[t]{0.42\textwidth}
		\centering
		\begin{tabular}{lrrr}
            \multicolumn{4}{@{}c}{ \textbf{Dependence on emissions cap $a$}  }\\
			\toprule
			$a$ & $0.54$ & $0.49$ & $0.39$ \\
			\midrule
$\mathrm{EE}$ & 8.41 & 9.48 & 12.17 \\
$\mu_{\mathbf{P}_t}$ & 71.30 & 75.02 & 82.48 \\
\midrule
$\eta_{a}(\mu_{\mathbf{P}_t})$ & -0.56 & -0.49 & -0.35 \\
$\eta_{a}(\pmb{\alpha}^{c_1}(0))$ & -0.78 & -0.66 & -0.47 \\
$\eta_{a}(\pmb{\alpha}^{c_1}(1))$ & -0.79 & -0.67 & -0.47 \\
$\eta_{a}(\pmb{\alpha}^{c_2}(0))$ & -0.87 & -0.73 & -0.51 \\
$\eta_{a}(\pmb{\alpha}^{c_2}(1))$ & -0.88 & -0.74 & -0.51 \\
$\eta_{a}(\pmb{\alpha}^{c_3}(0))$ & -0.97 & -0.81 & -0.55 \\
$\eta_{a}(\pmb{\alpha}^{c_3}(1))$ & -0.98 & -0.81 & -0.56 \\
$\eta_{a}(\pmb{\alpha}^{c_4}(0))$ & -1.11 & -0.91 & -0.61 \\
$\eta_{a}(\pmb{\alpha}^{c_4}(1))$ & -1.11 & -0.91 & -0.61 \\
$\eta_{a}(\pmb{\alpha}^{d_1}(0))$ & -1.29 & -1.04 & -0.68 \\
$\eta_{a}(\pmb{\alpha}^{d_1}(1))$ & -1.28 & -1.04 & -0.68 \\
$\eta_{a}(\pmb{\alpha}^{d_2}(0))$ & -1.53 & -1.22 & -0.78 \\
$\eta_{a}(\pmb{\alpha}^{d_2}(1))$ & -1.52 & -1.21 & -0.77 \\
$\eta_{a}(\pmb{\alpha}^{d_3}(0))$ & -1.89 & -1.46 & -0.90 \\
$\eta_{a}(\pmb{\alpha}^{d_3}(1))$ & -1.86 & -1.44 & -0.88 \\
$\eta_{a}(\pmb{\alpha}^{d_4}(0))$ & -2.47 & -1.82 & -1.06 \\
$\eta_{a}(\pmb{\alpha}^{d_4}(1))$ & -2.42 & -1.79 & -1.04 \\
			\bottomrule
		\end{tabular}
	\end{minipage}
\end{table}

Table~\ref{tab:lambda_theta_sidebyside} shows that raising the penalty rate $\lambda$ or tightening the emissions cap (i.e. lowering $a$) both increase the expected allowance price $\mu_{\mathbf{P}_t}$, consistent with \citet{Yu2020}.
The elasticity $\eta_\lambda(\mu_{\mathbf{P}_t})$ of allowance price with respect to $\lambda$ is nonlinear.
It is large when the penalty $\lambda$ is low and decreases as $\lambda$ rises, reflecting diminishing marginal effects and the saturation of firms' abatement responses. 
This saturation mechanism is driven by the progressive decline in expected excess emissions (EE) under more stringent policies.

When the penalty rate $\lambda$ is low, firms tend to emit beyond their allocated caps, making their abatement efforts highly sensitive to changes in $\lambda$---as reflected in the relatively high values of the elasticity $\eta_\lambda(\pmb{\alpha}^i(t))$ of firms' abatement efforts in this regime.
By contrast, when $\lambda$ is already severe, firms are effectively disciplined to stay within the cap, so further increases in $\lambda$ trigger only minor adjustments in abatement, producing a muted elasticity of allowance prices. 
This nonlinear pattern supports the results of \citet{Seifert2008} and contrasts with the linear price-penalty relationship reported by \citet{huang2022}.

The right panel of Table~\ref{tab:lambda_theta_sidebyside} shows that the elasticity $\eta_a(\mu_{\mathbf{P}_t})$ of allowance price  with respect to the emissions cap $a$ is negative, nonlinear, and of larger absolute magnitude than the elasticity with respect to $\lambda$, especially at higher penalty levels. 
This suggests that adjustments to the cap exert a stronger influence on allowance prices than comparable changes in the penalty rate once $\lambda$ is already high. 
The negative relationship arises because a higher cap $a$ corresponds to a less stringent environmental policy: firms face weaker incentives to abate and are willing to emit more. 
The lower emissions cap $a$ results in higher allowance prices, since the overall reduction in abatement effort raises aggregate emissions and intensifies demand for permits. 
This mechanism helps to explain the strong nonlinear sensitivity observed and is consistent with the findings of \citet{jiang2021}.

The absolute values of the elasticities of allowance prices $\mu_{\mathbf{P}_t}$  and abatement efforts $\pmb{\alpha}^i(t)$ with respect to $a$ exceed those with respect to $\lambda$, particularly at higher values of $\lambda$.
This indicates that the emissions cap $a$ is a stronger determinant of market prices and the decisions of individual firms to green themselves.

\begin{figure}[htp]
	\centering
	\begin{subfigure}[t]{0.49\textwidth}
		\centering
		\includegraphics[width=\textwidth]{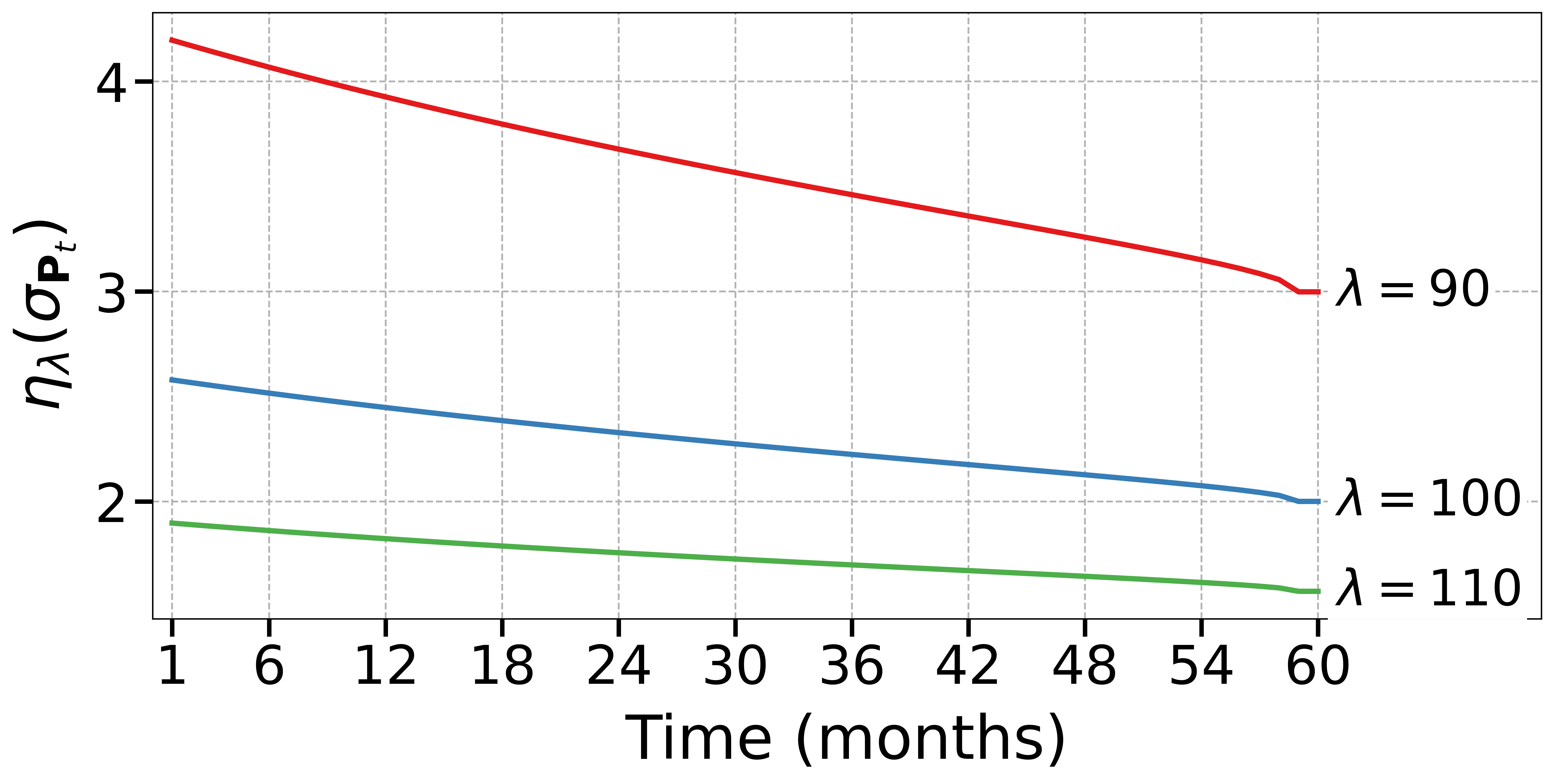}
		\caption{Dependence on penalty $\lambda$.}
		\label{fig:std_lambda}
	\end{subfigure}
	\hfill
	\begin{subfigure}[t]{0.49\textwidth}
		\centering
		\includegraphics[width=\textwidth]{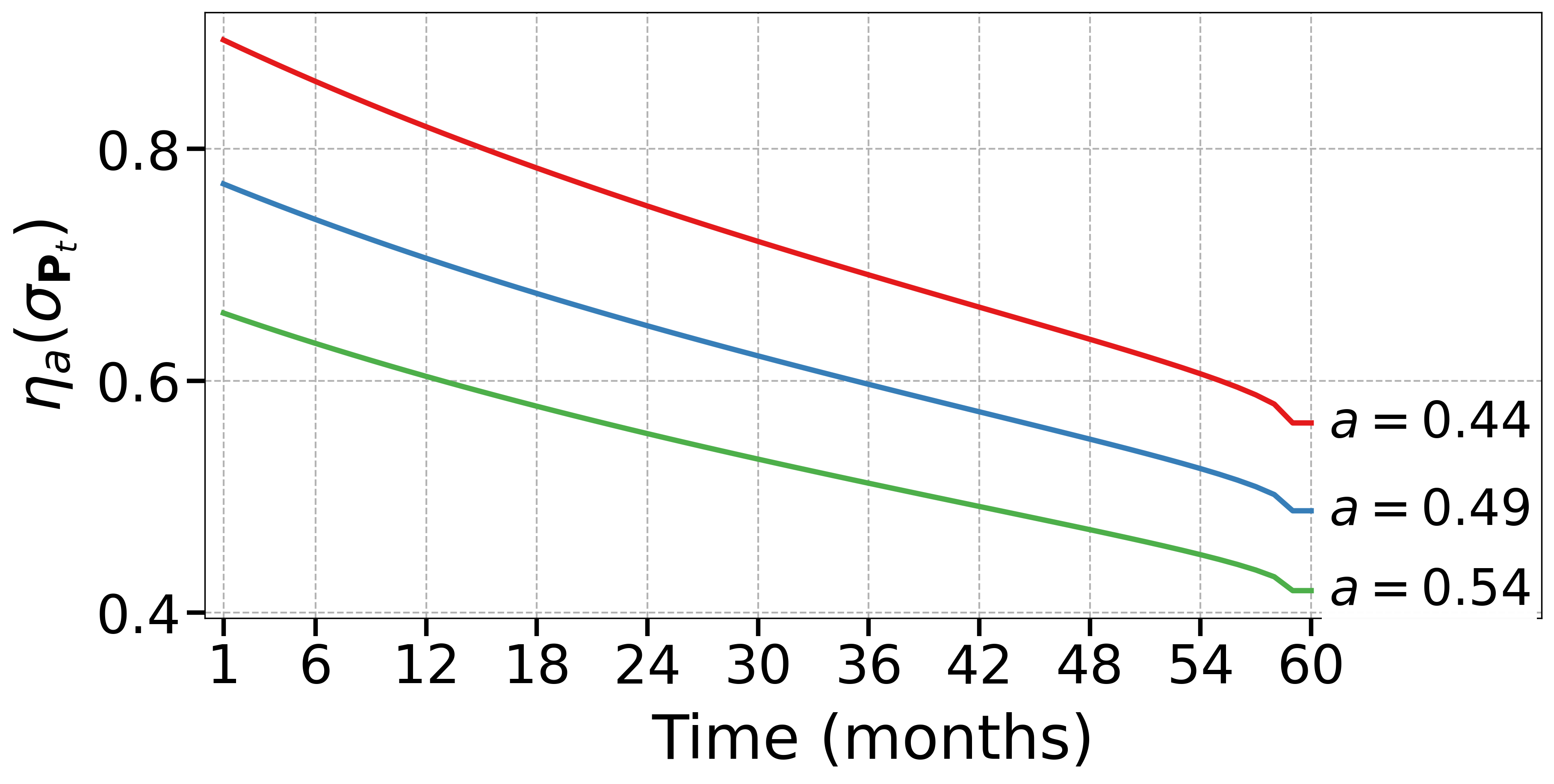}
		\caption{Dependence on  cap $a$.}
		\label{fig:std_theta}
	\end{subfigure}
	\caption{
			Elasticity of the standard deviation of the allowance price $\sigma_{\mathbf{P}_t}$ with respect to the penalty $\lambda$ (in €) and the emissions cap $a$, shown over time.
			}
	\label{fig:std_lambda_and_theta}
\end{figure}

Figure~\ref{fig:std_lambda_and_theta} depicts how regulatory parameters influence the volatility of allowance prices, measured by its standard deviation  $\sigma_{\mathbf{P}_t}$. 
The elasticity $\eta_\lambda\big(\sigma_{\mathbf{P}_t}\big)$  of $\sigma_{\mathbf{P}_t}$ with respect to the penalty rate $\lambda$ is positive, indicating that higher penalty rates  tend to increase price volatility. 
This effect is especially pronounced at lower penalty levels, suggesting that modest penalties have a stronger volatility-inducing impact, consistent with the findings of \cite{Seifert2008}. 
In comparison, the elasticity $\eta_a\big(\sigma_{\mathbf{P}_t}\big)$ of $\sigma_{\mathbf{P}_t}$ with respect to the emissions cap $a$ is also positive but of smaller magnitude, indicating that variations in $a$ exert a less substantial influence on price volatility than do changes in the penalty rate.
Notably, the elasticity of $\sigma_{\mathbf{P}_t}$ increases as the emissions cap becomes more stringent (lower $a$), highlighting that price-based measures (such as increasing free allowance supply when prices rise) tend to stabilize the market. 
This observation aligns with the recent works of \citet{Heijmans2023}.
Overall, the influence of $\lambda$ on $\sigma_{\mathbf{P}_t}$ is more substantial, particularly at lower penalty levels, than that of $a$, suggesting that cap stringency is more effective at stabilizing prices than increasing penalty rates. 
Furthermore, the sensitivity of $\sigma_{\mathbf{P}_t}$ to both $\lambda$ and $a$ declines as the market approaches the compliance deadline $T$, reflecting a diminishing regulatory influence over time.

\subsection{Dependence on the Abatement Cost Parameters}\label{subsec:abatementcost}

The abatement cost parameters---whether linear coefficients $\mathrm{k}^i$ or quadratic coefficients $\gamma^i$---exhibit substantial variability in the literature, largely reflecting the heterogeneity in firms' technological options for reducing carbon emissions, as highlighted by \citet{Rekker2023}.
A widely adopted benchmark for assessing optimal climate policy transitions is the DICE model developed by \citet{nordhaus2024}. 
Across successive updates of the model, the backstop price, corresponding to the abatement cost coefficients in our setup, has increased consistently, more than doubling in its most recent version.
This uncertainty motivates our sensitivity analysis, undertaken for parameters $\mathrm{k}^i$ (linear) and $\gamma^i$ (quadratic).

In a perfectly competitive market, equilibrium allowance prices align with marginal abatement costs, as noted by \citet{Wu2023} and \citet{Rekker2023}. 
This relationship implies that higher abatement cost coefficients ($\mathrm{k}^i$ or $\gamma^i$) result in an increase in the expected allowance price $\mu_{\mathbf{P}_t}$.
This is supported by the positive elasticities reported in Table~\ref{tab:elasticity_abatement}, and also consistent with the findings of \citet{Seifert2008}, \citet{Yu2020} and \citet{huang2022}.
Comparatively, $\mu_{\mathbf{P}_t}$ exhibits larger elasticities for firms with higher linear abatement cost coefficients $\mathrm{k}^i$ than for lower-cost firms.
Conversely, $\mu_{\mathbf{P}_t}$ exhibits larger elasticities for the quadratic coefficients $\gamma^i$ of lower-cost firms than for those of higher-cost firms, as shown in Table~\ref{tab:elasticity_abatement}.

An analysis of firms' optimal abatement plans $\pmb{\alpha}^i(t)$ in Table~\ref{tab:elasticity_abatement} reveals that own-cost elasticities are uniformly negative, aligning with theoretical expectations that higher linear ($\mathrm{k}^i$) or quadratic ($\gamma^i$) cost coefficients lead to diminished abatement efforts.
This reduction is markedly stronger for quadratic cost parameters ($\gamma^i$) than for linear ones ($\mathrm{k}^i$). 
These pronounced elasticities underscore how cost heterogeneity drives behavioral divergence and facilitates inter-firm leakage.  
Cross-partial elasticities, which capture firm's $i$ abatement response $\pmb{\alpha}^i(t)$ to changes in other firms' cost parameters, are generally modest yet consistently positive. 
These elasticities capture price-mediated strategic interactions: when firm $j$ reduces abatement in response to higher costs, the resulting increase in equilibrium prices incentivizes other firms to expand their abatement efforts, effectively substituting for the reduction by firm $j$.
 Notably, firms' abatement responses are most pronounced to changes in the linear cost coefficient ($\mathrm{k}^j$) of high-cost firms $j$, whereas variations in the quadratic coefficient ($\gamma^j$) elicit stronger responses among low-cost firms.

\begin{table}[htp!]
\centering
\rotatebox{90}{%
\begin{minipage}{\textheight}
\centering
\caption{Elasticities of the expected allowance price $\mu_{\mathbf{P}_t}$ and the optimal abatement plans $\pmb{\alpha}^i(t)$ with respect
to the linear cost coefficients $\mathrm{k}^i$ (top panel) and the quadratic cost coefficients $\gamma^i$ (bottom panel).}
\label{tab:elasticity_abatement}
\small
\setlength{\tabcolsep}{4pt}
\renewcommand{\arraystretch}{1.1}
\begin{tabular}{lrrrrrrrrrrrrrrrrr}
\toprule
\multicolumn{18}{@{}c}{\textit{Elasticities with respect to abatement cost parameters}} \\
\cmidrule(lr){2-18}
 Parameter & $\mu_{\mathbf{P}_t}$ & $\pmb{\alpha}^{c_1}(0)$ & $\pmb{\alpha}^{c_1}(1)$ & $\pmb{\alpha}^{c_2}(0)$ & $\pmb{\alpha}^{c_2}(1)$ & $\pmb{\alpha}^{c_3}(0)$ & $\pmb{\alpha}^{c_3}(1)$ & $\pmb{\alpha}^{c_4}(0)$ & $\pmb{\alpha}^{c_4}(1)$ & $\pmb{\alpha}^{d_1}(0)$ & $\pmb{\alpha}^{d_1}(1)$ & $\pmb{\alpha}^{d_2}(0)$ & $\pmb{\alpha}^{d_2}(1)$ & $\pmb{\alpha}^{d_3}(0)$ & $\pmb{\alpha}^{d_3}(1)$ & $\pmb{\alpha}^{d_4}(0)$ & $\pmb{\alpha}^{d_4}(1)$ \\
\midrule
\multicolumn{18}{@{}l}{\textit{Linear coefficients}} \\
$\mathrm{k}^{c_1}$ &  0.04 & -0.31 & -0.31 &  0.05 &  0.05 &  0.06 &  0.06 &  0.07 &  0.07 &  0.08 &  0.08 &  0.09 &  0.09 &  0.11 &  0.11 &  0.13 &  0.13 \\
$\mathrm{k}^{c_2}$ &  0.04 &  0.06 &  0.06 & -0.43 & -0.43 &  0.07 &  0.07 &  0.08 &  0.08 &  0.09 &  0.09 &  0.11 &  0.11 &  0.13 &  0.13 &  0.16 &  0.16 \\
$\mathrm{k}^{c_3}$ &  0.05 &  0.07 &  0.07 &  0.08 &  0.08 & -0.58 & -0.57 &  0.10 &  0.09 &  0.11 &  0.11 &  0.13 &  0.13 &  0.15 &  0.15 &  0.19 &  0.19 \\
$\mathrm{k}^{c_4}$ &  0.06 &  0.08 &  0.08 &  0.09 &  0.09 &  0.10 &  0.10 & -0.76 & -0.75 &  0.13 &  0.12 &  0.15 &  0.14 &  0.18 &  0.17 &  0.22 &  0.21 \\
$\mathrm{k}^{d_1}$ &  0.07 &  0.09 &  0.09 &  0.10 &  0.10 &  0.11 &  0.11 &  0.12 &  0.12 & -1.00 & -0.99 &  0.16 &  0.16 &  0.20 &  0.19 &  0.25 &  0.24 \\
$\mathrm{k}^{d_2}$ &  0.07 &  0.10 &  0.10 &  0.11 &  0.11 &  0.12 &  0.12 &  0.14 &  0.13 &  0.16 &  0.15 & -1.32 & -1.30 &  0.22 &  0.21 &  0.27 &  0.26 \\
$\mathrm{k}^{d_3}$ &  0.08 &  0.11 &  0.11 &  0.12 &  0.12 &  0.13 &  0.13 &  0.15 &  0.15 &  0.17 &  0.17 &  0.20 &  0.19 & -1.76 & -1.73 &  0.30 &  0.28 \\
$\mathrm{k}^{d_4}$ &  0.09 &  0.12 &  0.12 &  0.13 &  0.13 &  0.14 &  0.14 &  0.16 &  0.16 &  0.18 &  0.18 &  0.21 &  0.21 &  0.26 &  0.24 & -2.43 & -2.38 \\
\midrule
\multicolumn{18}{@{}l}{\textit{Quadratic coefficients}} \\
$\gamma^{c_1}$ &  0.10 & -0.87 & -0.86 &  0.15 &  0.14 &  0.16 &  0.16 &  0.18 &  0.18 &  0.21 &  0.21 &  0.24 &  0.24 &  0.29 &  0.29 &  0.37 &  0.37 \\
$\gamma^{c_2}$ &  0.09 &  0.12 &  0.12 & -0.87 & -0.86 &  0.15 &  0.14 &  0.16 &  0.16 &  0.19 &  0.18 &  0.22 &  0.22 &  0.26 &  0.26 &  0.33 &  0.32 \\
$\gamma^{c_3}$ &  0.08 &  0.10 &  0.10 &  0.12 &  0.11 & -0.87 & -0.86 &  0.14 &  0.14 &  0.16 &  0.16 &  0.19 &  0.19 &  0.23 &  0.23 &  0.29 &  0.28 \\
$\gamma^{c_4}$ &  0.07 &  0.09 &  0.09 &  0.10 &  0.10 &  0.11 &  0.11 & -0.87 & -0.86 &  0.14 &  0.14 &  0.17 &  0.16 &  0.20 &  0.20 &  0.25 &  0.24 \\
$\gamma^{d_1}$ &  0.06 &  0.08 &  0.08 &  0.09 &  0.09 &  0.10 &  0.10 &  0.11 &  0.11 & -0.88 & -0.87 &  0.14 &  0.14 &  0.17 &  0.17 &  0.22 &  0.21 \\
$\gamma^{d_2}$ &  0.05 &  0.07 &  0.07 &  0.07 &  0.07 &  0.08 &  0.08 &  0.09 &  0.09 &  0.10 &  0.10 & -0.88 & -0.87 &  0.15 &  0.14 &  0.18 &  0.17 \\
$\gamma^{d_3}$ &  0.04 &  0.05 &  0.06 &  0.06 &  0.06 &  0.07 &  0.07 &  0.07 &  0.07 &  0.09 &  0.08 &  0.10 &  0.10 & -0.88 & -0.87 &  0.15 &  0.14 \\
$\gamma^{d_4}$ &  0.03 &  0.04 &  0.04 &  0.05 &  0.05 &  0.05 &  0.05 &  0.06 &  0.06 &  0.07 &  0.07 &  0.08 &  0.08 &  0.09 &  0.09 & -0.88 & -0.88 \\
\bottomrule
\end{tabular}
\end{minipage}%
}
\end{table}

Figure~\ref{fig:std_abat} illustrates the elasticities of the standard deviation of allowance prices, $\sigma_{\mathbf{P}_t}$, with respect to abatement cost parameters. 
These elasticities are uniformly negative, indicating that increases in either $\mathrm{k}^i$ or $\gamma^i$ generally reduce price volatility. 
However, their magnitudes are relatively modest, suggesting that price volatility exhibits limited sensitivity to variations in abatement costs. 
Notably, the observed positive elasticity of $\sigma_{\mathbf{P}_t}$ with respect to $\gamma^i$ contrasts with the findings of \citet{Seifert2008}, though it is important to mention that \citet{Seifert2008} uses quadratic cost (not linear quadratic) and analyze the problem from a social planner's perspective. 
The figure further reveals that $\sigma_{\mathbf{P}_t}$ responds more strongly to changes in the linear cost coefficients of higher-cost firms, while volatility is more sensitive to variations in the quadratic coefficients of lower-cost firms. 
In both parameters, the sensitivity of $\sigma_{\mathbf{P}_t}$ increases as the market approaches the compliance deadline $T$.

\begin{figure}[htp]
	\centering
	\begin{subfigure}[t]{0.49\textwidth}
		\centering
		\includegraphics[width=\textwidth]{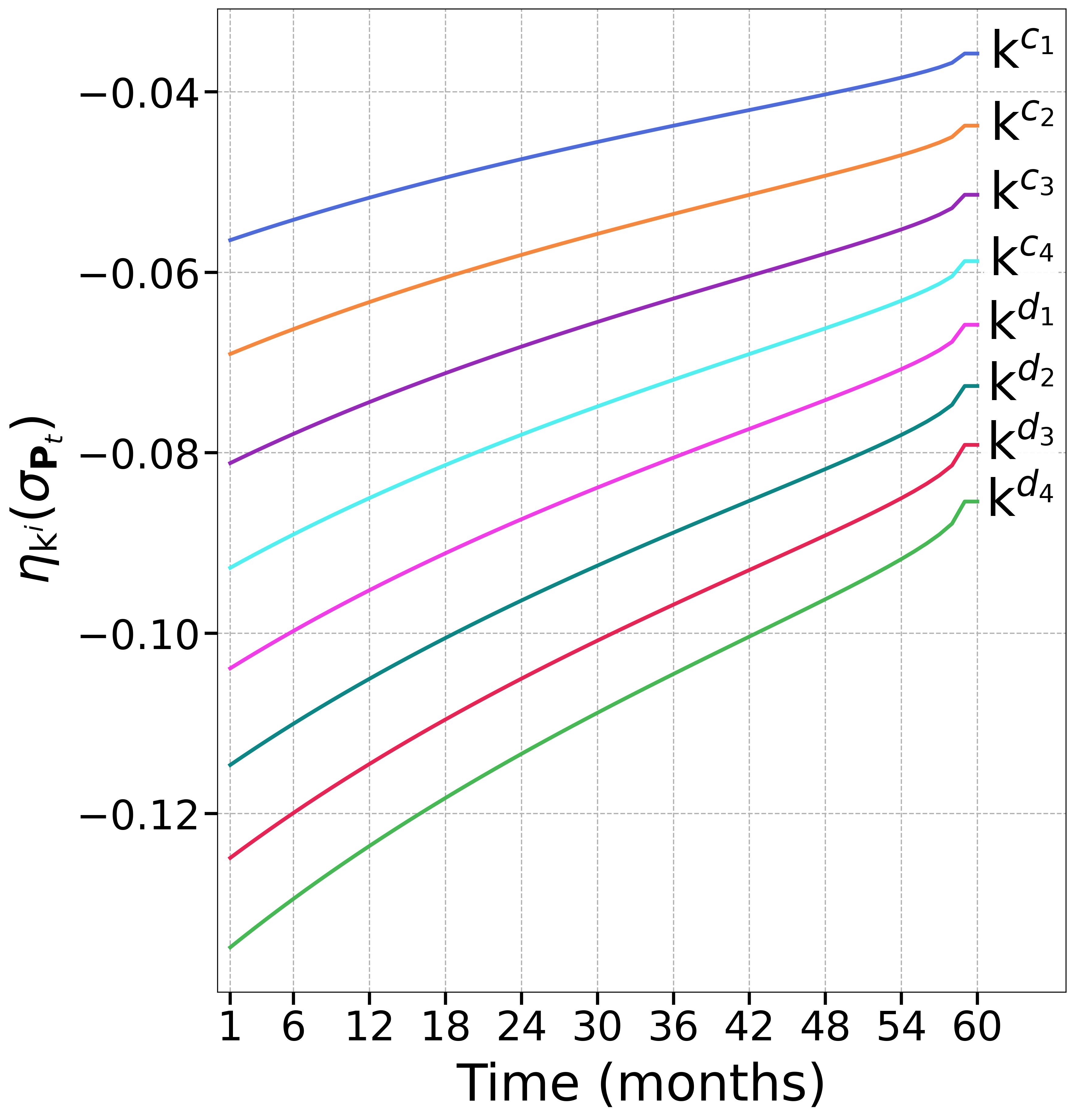}
		\caption{Dependence on $\mathrm{k}^i$.}
		\label{fig:std_linabat}
	\end{subfigure}
		\hfill
	\begin{subfigure}[t]{0.49\textwidth}
		\centering
		\includegraphics[width=\textwidth]{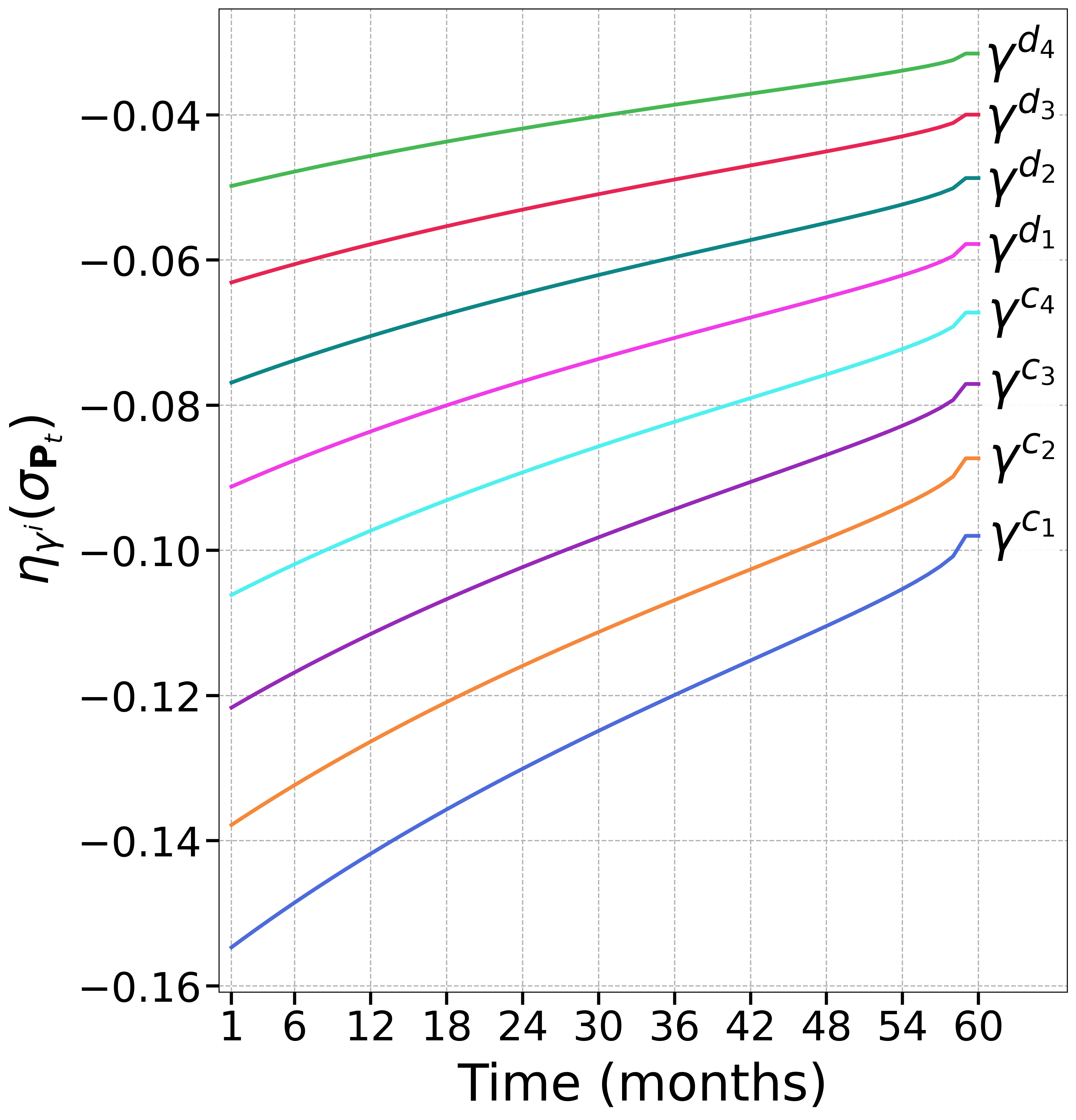}
		\caption{Dependence on $\gamma^i$.}
		\label{fig:std_quadabat}
	\end{subfigure}
	\caption{
	Elasticity of the standard deviation of allowance price  $\sigma_{\mathbf{P}_t}$ with respect to linear abatement cost coefficients $\mathrm{k}^i$ (left panel) and quadratic abatement cost coefficients $\gamma^i$ (right panel). The value of the parameters are given in Table~\ref{tab:parameters}.
	}
	\label{fig:std_abat}
\end{figure}

Finally, we examine how expected excess emissions $\mathrm{EE}$ vary in response to changes in abatement cost parameters. 
Generally, one would expect $\mathrm{EE}$ to increase as costs rise, since higher costs tend to discourage abatement efforts.
However, exceptions are observed for the linear cost coefficients of lower-cost firms. 
For example, $\mathrm{EE}$ decreases as the linear cost coefficient $\mathrm{k}^{c_1}$ increases, as shown in left panel of Table~\ref{tab:kc1_kd4_sensitivity}. 
This counterintuitive behavior can be explained by two factors: (i) the relatively limited abatement adjustment of firm $c_1$ to changes in its own linear cost $\mathrm{k}^{c_1}$, and (ii) more cautious abatement responses from other firms as firm's $c_1$ linear cost rises.
These dynamics are reflected in the elasticity patterns of the abatement plans $\pmb{\alpha}$ presented in Table~\ref{tab:elasticity_abatement}.

\begin{table}[htp]
\caption{{The expected excess emissions $\mathrm{EE}$ (in $10^6$ tons) and the expected allowance price \
$\mu_{\mathbf{P}_t}$ (€) for different values of the linear abatement cost coefficients $\mathrm{k}^{c_1}$ and $\mathrm{k}^{d_4}$ (€).}}
\label{tab:kc1_kd4_sensitivity}
\centering

\begin{minipage}{0.48\textwidth}
\centering
\small
\begin{tabular}{lrrrrr}
\multicolumn{6}{c}{\textbf{Dependence on $\mathrm{k}^{c_1}$}} \\
\toprule
$\mathrm{k}^{c_1}$ & $15.0$ & $17.5$ & $20.0$ & $22.5$ & $25.0$ \\
\midrule
$\mathrm{EE}$ & 10.08 & 9.77 & 9.48 & 9.21 & 8.96 \\
$\mu_{\mathbf{P}_t}$ & 74.35 & 74.69 & 75.02 & 75.36 & 75.69 \\
\bottomrule
\end{tabular}
\label{tab:kc1_sensitivity}
\end{minipage}
\hfill
\begin{minipage}{0.48\textwidth}
\centering
\small
\begin{tabular}{lrrrrr}
\multicolumn{6}{c}{\textbf{Dependence on $\mathrm{k}^{d_4}$}} \\
\toprule
$\mathrm{k}^{d_4}$ & $45.0$ & $47.5$ & $50.0$ & $52.5$ & $55.0$ \\
\midrule
$\mathrm{EE}$ & 8.12 & 8.41 & 8.74 & 9.09 & 9.48 \\
$\mu_{\mathbf{P}_t}$ & 73.86 & 74.15 & 74.44 & 74.73 & 75.02 \\
\bottomrule
\end{tabular}
\label{tab:kd4_sensitivity}
\end{minipage}

\end{table}

\subsection{Dependence on the  Emissions Parameters}\label{sec:emis}
This section examines the sensitivity of market outcomes to variations in firm-level emission characteristics, focusing specifically on the effects of changes in the mean ($\mu$) and standard deviation ($\sigma$) of business-as-usual (BAU) emissions, as well as the correlation coefficient ($\rho$) of emissions across firms.

\begin{table}[htp!]
    \centering
    \caption{The expected excess emissions $\mathrm{EE}$ (in $10^6$ tons), the expected allowance price\
    $\mu_{\mathbf{P}_t}$ (€), and the elasticities of $\mu_{\mathbf{P}_t}$ and abatement plans $\pmb{\alpha}^i(t)$ with respect to firm-level  mean\
    $\mu$ (in $10^7$ tons) (left panel) and standard deviation $\sigma$ (in $10^7$ tons) (right panel) of emissions.}
    \label{tab:mu_sigma_sidebyside}
    \begin{minipage}[t]{0.48\textwidth}
        \centering
        \begin{tabular}{lrrr}
            \multicolumn{4}{c}{\textbf{Dependence on $\mu$}} \\
           \toprule
            $\mu$ & $2.55$ & $2.71$ & $2.86$ \\
            \midrule
$\mathrm{EE}$ & 9.37 & 9.48 & 9.61 \\
$\mu_{\mathbf{P}_t}$ & 73.14 & 75.04 & 76.85 \\
\midrule
$\eta_{\mu}(\mu_{\mathbf{P}_t})$ & 0.41 & 0.43 & 0.45 \\
$\eta_{\mu}(\pmb{\alpha}^{c_1}(0))$ & -0.43 & -0.41 & -0.39 \\
$\eta_{\mu}(\pmb{\alpha}^{c_1}(1))$ & -0.25 & -0.25 & -0.24 \\
$\eta_{\mu}(\pmb{\alpha}^{c_2}(0))$ & -0.37 & -0.35 & -0.33 \\
$\eta_{\mu}(\pmb{\alpha}^{c_2}(1))$ & -0.19 & -0.19 & -0.19 \\
$\eta_{\mu}(\pmb{\alpha}^{c_3}(0))$ & -0.30 & -0.28 & -0.26 \\
$\eta_{\mu}(\pmb{\alpha}^{c_3}(1))$ & -0.13 & -0.12 & -0.12 \\
$\eta_{\mu}(\pmb{\alpha}^{c_4}(0))$ & -0.21 & -0.19 & -0.18 \\
$\eta_{\mu}(\pmb{\alpha}^{c_4}(1))$ & -0.04 & -0.04 & -0.04 \\
$\eta_{\mu}(\pmb{\alpha}^{d_1}(0))$ & -0.09 & -0.07 & -0.06 \\
$\eta_{\mu}(\pmb{\alpha}^{d_1}(1))$ & 0.07 & 0.07 & 0.07 \\
$\eta_{\mu}(\pmb{\alpha}^{d_2}(0))$ & 0.07 & 0.08 & 0.08 \\
$\eta_{\mu}(\pmb{\alpha}^{d_2}(1))$ & 0.23 & 0.22 & 0.21 \\
$\eta_{\mu}(\pmb{\alpha}^{d_3}(0))$ & 0.30 & 0.29 & 0.29 \\
$\eta_{\mu}(\pmb{\alpha}^{d_3}(1))$ & 0.44 & 0.42 & 0.40 \\
$\eta_{\mu}(\pmb{\alpha}^{d_4}(0))$ & 0.66 & 0.62 & 0.58 \\
$\eta_{\mu}(\pmb{\alpha}^{d_4}(1))$ & 0.78 & 0.72 & 0.68 \\
            \bottomrule
        \end{tabular}
    \end{minipage}
    \hfill
    \begin{minipage}[t]{0.48\textwidth}
        \centering
        \begin{tabular}{lrrr}
            \multicolumn{4}{c}{\textbf{Dependence on $\sigma$}} \\
            \toprule
            $\sigma$ & $0.61$ & $0.79$ & $0.96$ \\
            \midrule
$\mathrm{EE}$ & 7.34 & 9.53 & 11.65 \\
$\mu_{\mathbf{P}_t}$ & 73.90 & 75.05 & 76.40 \\
\midrule
$\eta_{\sigma}(\mu_{\mathbf{P}_t})$ & 0.05 & 0.08 & 0.11 \\
$\eta_{\sigma}(\pmb{\alpha}^{c_1}(0))$ & 0.06 & 0.10 & 0.15 \\
$\eta_{\sigma}(\pmb{\alpha}^{c_1}(1))$ & -0.04 & -0.05 & -0.07 \\
$\eta_{\sigma}(\pmb{\alpha}^{c_2}(0))$ & 0.07 & 0.11 & 0.16 \\
$\eta_{\sigma}(\pmb{\alpha}^{c_2}(1))$ & -0.03 & -0.04 & -0.06 \\
$\eta_{\sigma}(\pmb{\alpha}^{c_3}(0))$ & 0.08 & 0.13 & 0.18 \\
$\eta_{\sigma}(\pmb{\alpha}^{c_3}(1))$ & -0.02 & -0.03 & -0.04 \\
$\eta_{\sigma}(\pmb{\alpha}^{c_4}(0))$ & 0.09 & 0.14 & 0.20 \\
$\eta_{\sigma}(\pmb{\alpha}^{c_4}(1))$ & -0.01 & -0.01 & -0.02 \\
$\eta_{\sigma}(\pmb{\alpha}^{d_1}(0))$ & 0.10 & 0.16 & 0.23 \\
$\eta_{\sigma}(\pmb{\alpha}^{d_1}(1))$ & 0.01 & 0.01 & 0.01 \\
$\eta_{\sigma}(\pmb{\alpha}^{d_2}(0))$ & 0.12 & 0.19 & 0.27 \\
$\eta_{\sigma}(\pmb{\alpha}^{d_2}(1))$ & 0.03 & 0.04 & 0.05 \\
$\eta_{\sigma}(\pmb{\alpha}^{d_3}(0))$ & 0.14 & 0.23 & 0.32 \\
$\eta_{\sigma}(\pmb{\alpha}^{d_3}(1))$ & 0.05 & 0.08 & 0.10 \\
$\eta_{\sigma}(\pmb{\alpha}^{d_4}(0))$ & 0.18 & 0.28 & 0.39 \\
$\eta_{\sigma}(\pmb{\alpha}^{d_4}(1))$ & 0.09 & 0.14 & 0.18 \\
            \bottomrule
        \end{tabular}
    \end{minipage}
\end{table}

Table~\ref{tab:mu_sigma_sidebyside} reports the elasticities of the expected allowance prices $\mu_{\mathbf{P}_t}$ with respect to the mean ($\mu$) and standard deviation ($\sigma$) of firms' BAU emissions.
The results show that both elasticities, $\eta_\mu\big(\mu_{\mathbf{P}_t}\big)$ and $\eta_\sigma\big(\mu_{\mathbf{P}_t}\big)$, of the allowance price with respect to $\mu$ and $\sigma$ are positive, indicating that increases in either the mean level or the volatility of emissions lead to higher expected allowance prices.
Moreover, $\mu_{\mathbf{P}_t}$  exhibits higher sensitivity to changes in $\mu$ than to changes in $\sigma$.
The observed monotonic relationship between allowance prices and the mean of firms' emissions is consistent with the findings of  \citet{Seifert2008} and \citet{guo2019}.
Conversely, \citet{Seifert2008} noted that allowance prices may either increase or decrease in response to variations in $\sigma$.
  
As illustrated in the left panel of Table~\ref{tab:mu_sigma_sidebyside}, firm-level abatement responses to changes in $\mu$ vary according to firms' abatement costs.
Firms with relatively low abatement costs (e.g. $c_1$--$c_4$ and $d_1$) exhibit negative abatement elasticities with respect to $\mu$, meaning they reduce abatement as baseline emissions increase. 
Conversely, firms with higher abatement costs show positive elasticities, implying they bear a larger share of the adjustment burden. 
This asymmetry suggests that as aggregate emissions rise, the regulatory pressure disproportionately shifts toward firms with more expensive abatement options.

The right panel of Table~\ref{tab:mu_sigma_sidebyside} also shows that the elasticities of abatement plans $\pmb{\alpha}$  with respect to $\sigma$ exhibit  temporal heterogeneity. 
Firms' abatement decisions at initial time $0$ are positively sensitive in $\sigma$, whereas after time $0$ their responses decline or even become negative---particularly for firms with lower abatement costs. 
This pattern points to a strategic reallocation of abatement efforts toward earlier periods in the face of heightened uncertainty, likely reflecting a precautionary motive to hedge against future price volatility.
This pattern is also observed in the recent findings of \cite{guo2019}.

Table~\ref{tab:mu_sigma_sidebyside} shows that expected excess emissions, $\mathrm{EE}$, rise with both the mean ($\mu$) and standard deviation ($\sigma$) of firms' BAU emissions, since higher values of these parameters results higher expected emissions, see~\eqref{eq:g1}.

\begin{table}[htp!]
     \caption{The expected excess emissions $\mathrm{EE}$ (in $10^6$ tons), the expected allowance price\
 $\mu_{\mathbf{P}_t}$ (€), and the elasticities of $\mu_{\mathbf{P}_t}$ and abatement plans $\pmb{\alpha}^i(t)$\
 with respect to the correlation coefficient ($\rho$) between firms’ emissions.}
 \label{tab:rho_table}
	\centering
		\begin{tabular}{lrrrrr}
			
            \toprule
			$\rho$ & $0.71$ & $0.78$ & $0.85$ & $0.92$ & $0.99$ \\
			
           \midrule
$\mathrm{EE}$ & 13.11 & 11.44 & 9.48 & 6.95 & 2.48 \\
$\mu_{\mathbf{P}_t}$ & 75.02 & 75.02 & 75.02 & 75.02 & 74.95 \\
\midrule
$\eta_{\rho}(\mu_{\mathbf{P}_t})$ & -0.03 & -0.03 & -0.03 & -0.03 & -0.11 \\
$\eta_{\rho}(\pmb{\alpha}^{c_1}(0))$ & -0.04 & -0.04 & -0.04 & -0.05 & -0.14 \\
$\eta_{\rho}(\pmb{\alpha}^{c_1}(1))$ & 0.03 & 0.04 & 0.04 & 0.05 & 0.09 \\
$\eta_{\rho}(\pmb{\alpha}^{c_2}(0))$ & -0.05 & -0.05 & -0.05 & -0.05 & -0.16 \\
$\eta_{\rho}(\pmb{\alpha}^{c_2}(1))$ & 0.03 & 0.03 & 0.03 & 0.04 & 0.08 \\
$\eta_{\rho}(\pmb{\alpha}^{c_3}(0))$ & -0.05 & -0.05 & -0.05 & -0.06 & -0.18 \\
$\eta_{\rho}(\pmb{\alpha}^{c_3}(1))$ & 0.02 & 0.02 & 0.02 & 0.03 & 0.05 \\
$\eta_{\rho}(\pmb{\alpha}^{c_4}(0))$ & -0.06 & -0.06 & -0.06 & -0.06 & -0.20 \\
$\eta_{\rho}(\pmb{\alpha}^{c_4}(1))$ & 0.01 & 0.01 & 0.01 & 0.01 & 0.03 \\
$\eta_{\rho}(\pmb{\alpha}^{d_1}(0))$ & -0.07 & -0.07 & -0.07 & -0.07 & -0.23 \\
$\eta_{\rho}(\pmb{\alpha}^{d_1}(1))$ & -0.00 & -0.00 & -0.00 & -0.00 & -0.00 \\
$\eta_{\rho}(\pmb{\alpha}^{d_2}(0))$ & -0.08 & -0.08 & -0.08 & -0.09 & -0.26 \\
$\eta_{\rho}(\pmb{\alpha}^{d_2}(1))$ & -0.02 & -0.02 & -0.02 & -0.03 & -0.04 \\
$\eta_{\rho}(\pmb{\alpha}^{d_3}(0))$ & -0.09 & -0.10 & -0.10 & -0.10 & -0.32 \\
$\eta_{\rho}(\pmb{\alpha}^{d_3}(1))$ & -0.04 & -0.05 & -0.05 & -0.06 & -0.10 \\
$\eta_{\rho}(\pmb{\alpha}^{d_4}(0))$ & -0.12 & -0.12 & -0.12 & -0.13 & -0.40 \\
$\eta_{\rho}(\pmb{\alpha}^{d_4}(1))$ & -0.08 & -0.08 & -0.09 & -0.10 & -0.18 \\
			
    \bottomrule
		\end{tabular}
\end{table}

Table~\ref{tab:rho_table} presents the sensitivity results with respect to the correlation coefficient $\rho$ between firms' emissions. 
A higher correlation implies that when emissions rise in one firm, the other firm experiences  similar patterns.
Consequently, diversification or risk-sharing opportunities between sectors disappear.
As $\rho$ rises, firms are less able to offset each other's emission shocks, resulting in greater systemic risk.
The elasticities $\eta_\rho(\mu_{\mathbf{P}_t})$ of the expected allowance price $\mu_{\mathbf{P}_t}$ are negative and close to zero.
Because stronger correlation $\rho$ lowers both the mean $\mu_{\mathbf{P}_t}$ and volatility $\sigma_{\mathbf{P}_t}$ of allowance prices, aggregate emission fluctuations diminish, leading to a marked reduction in expected excess emissions $\mathrm{EE}$, as shown in the table.
The elasticities of firms' abatement plans $\pmb{\alpha}^i(t)$ with respect to $\rho$ are also small in magnitude across all firms and time periods.
Nonetheless, a distinct temporal pattern emerges: cleaner firms (with lower abatement costs) display positive elasticities in later periods---indicating that stronger correlation induces them to intensify abatement at later stages---whereas dirtier firms (with higher marginal abatement costs) show consistently negative elasticities, suggesting that higher correlation dampens their abatement effort over time.
This asymmetry highlights how heterogeneity in cost structures shapes firms' strategic responses to systemic emission shocks under increasing interdependence.

Figure~\ref{fig:std_emission} visualizes the elasticity of the standard deviation of allowance price $\sigma_{\mathbf{P}_t}$ with respect to the three emission parameters. 
Elasticities with respect to both $\mu$ and $\sigma$ are uniformly negative, indicating that increases in either the mean or volatility of emissions tend to reduce price volatility. 
However, the magnitudes of these effects are relatively small, especially for $\sigma$, suggesting that allowance price volatility exhibits limited sensitivity to variations in the volatility of individual firms' emissions. 
This observation contrasts with earlier findings by \citet{Seifert2008}, who reported a positive  relationship between $\sigma_{\mathbf{P}_t}$ and the standard deviation $\sigma$ of firms' emissions. 
In contrast, the elasticity with respect to the correlation coefficient $\rho$ is positive, albeit smaller in magnitude than those associated with $\mu$ and $\sigma$. 
This indicates that increased synchronization of emissions across firms modestly amplifies price volatility.
\begin{figure}[htp]
    \centering
    \includegraphics[width=\textwidth]{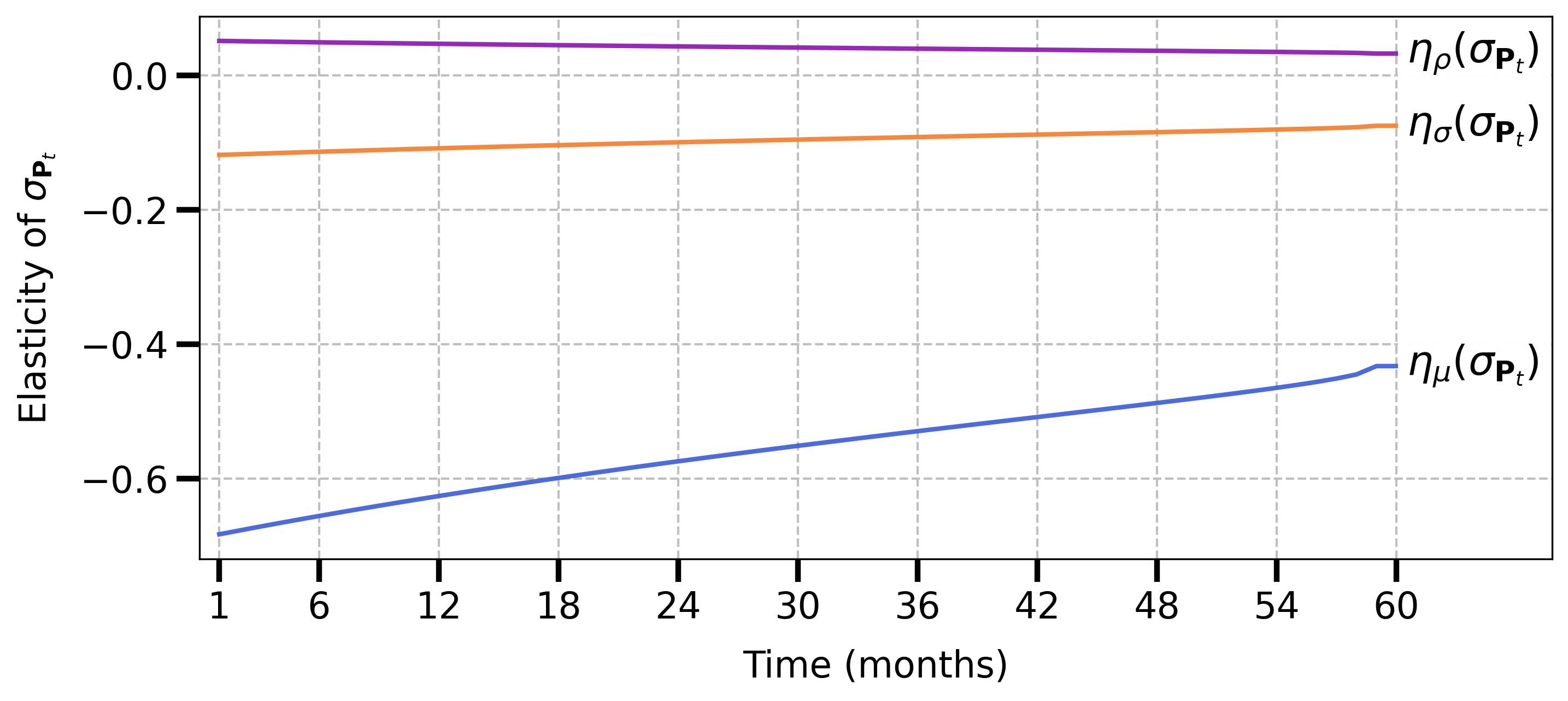}
    \caption{
			Elasticity of the standard deviation of allowance prices, $\sigma_{\mathbf{P}_t}$, with respect to the firm-level mean $\mu$, the standard deviation $\sigma$ of BAU emissions, and the correlation coefficient $\rho$ between firms' emissions, using parameter values specified in Table~\ref{tab:parameters}.
			}
    \label{fig:std_emission}
\end{figure}
Notably, $\sigma_{\mathbf{P}_t}$ appears somewhat more responsive to changes in $\mu$ than to variations in either $\sigma$ or $\rho$.
The elasticities with respect to both $\mu$ and $\sigma$ also decline in absolute value as the market approaches the compliance deadline $T$, indicating that the impact of emissions uncertainty on price volatility diminishes in the later stages of the trading period.

\section{Conclusion} \label{sec:conclusion}
In this paper, we examine how key regulatory, technological, and emissions-related factors shape allowance prices and firms' abatement strategies in a cap-and-trade system.
We extend the standard model within a Gaussian framework that endogenously determines allowance prices, trading strategies, and abatement decisions through a Radner equilibrium.
A semi-explicit formula is established for the allowance price and its variance, partially elucidating how parameters influence allowance pricing. 
Moreover, this formulation facilitates a sensitivity analysis based on the implicit function theorem that investigates (i) the elasticity of firms' average abatement efforts, (ii) the average elasticity of the allowance price, and (iii) the elasticity of price volatility with respect to fundamental model parameters.

The sensitivity analysis results in our model provide valuable insights for both regulatory authority and emitting firms, enhancing understanding of the extent to which firms undertake emission reduction initiatives and how changes in key input variables influence allowance prices. 
The elasticity measures quantify the degree to which variations in these parameters affect core market outcomes, thereby reflecting the interdependencies among their effects.
While our findings are derived within a Gaussian, risk-neutral framework, future research could extend the analysis by incorporating heterogeneous risk preferences among diverse market participants and exploring alternative baseline emissions distributions. 
We might also include additional risk factors, such as extreme events (droughts and floods) or tipping points, or features, such as financial intermediaries, richer allowance allocations rules, or
transaction costs. 
Such extensions would broaden the relevance and applicability of the results across varied market settings.

\appendix
\section{Proof of Proposition~\ref{prop:Fallocation}}\label{app:proof3}

Let $\mathbf{P}$, $\pmb{\alpha}$ and $(\mathbf{Q}^i)_{i }$ be a Radner equilibrium. 
If $\mathbf{P}$ is not a martingale, then $\mathbb{E}[\Delta \mathbf{P}_{t} 1_B]> 0$ or $<0$ holds for some $t \in 0\, .. \, T-1$ and  $B \in \mathfrak{F}_t$ with $\mathbb{P}[B]>0$.
If $\mathbb{E}[\Delta \mathbf{P}_{t} 1_B]> 0$, then for every $i$, for $\hat{Q}^i_s =\mathbf{Q}^i_s$, $s\neq t$,  and  $\hat{Q}^i_t =\mathbf{Q}^i_t + \mathds{1}_B$, we have
\begin{equation*}
	\mathbb{E} \big[ C^{i,\mathbf{P}}  ( \pmb{\alpha}^i,  \hat{Q}^i ) \big] = \mathbb{E} \big[ C^{i, \mathbf{P}}  ( \pmb{\alpha}^i,  \mathbf{Q}^i )  \big] -\mathbb{E}[\Delta \mathbf{P}_{t} 1_B] < \mathbb{E} \big[ C^{i, \mathbf{P}}  (  \pmb{\alpha}^i, \mathbf{Q}^i ) \big ],
\end{equation*}
which contradicts the optimality condition of firm $i$.
If instead $\mathbb{E}[\Delta \mathbf{P}_{t} 1_B]< 0$, using $\hat{Q}^i_s =\mathbf{Q}^i_s$, $s\neq t$, $\hat{Q}^i_t =\mathbf{Q}^i_t-\mathds{1}_B$ entails likewise a contradiction to the optimality condition of any firm $i$.
Hence, $\mathbf{P}$ is martingale.

\noindent
As $\mathbf{Q}^i$ is adapted, $\mathbf{Q}^i_T(\omega)$ then almost surely coincides  with some minimizer of 
\begin{equation}\label{eq:functionf}
	\mathbb{R} \ni	x \stackrel{g^{i,\pmb{\alpha}^i}_\omega}{\mapsto} x \mathbf{P}_T(\omega) + \lambda \big( E^{i, \pmb{\alpha}^i}_T(\omega) -A^i_T(\omega)-x \big)^+.
\end{equation}
By convexity of $g^{i,\pmb{\alpha}^i}_\omega$, $0 \in \partial g^{i,\pmb{\alpha}^i}_\omega\left(\mathbf{Q}^i_T(\omega) \right)=\big [\big(g^{i,\pmb{\alpha}^i}_\omega\big)'_- (\mathbf{Q}^i_T(\omega)), \big(g^{i,\pmb{\alpha}^i}_\omega\big)'_{+}(\mathbf{Q}^i_T(\omega)) \big ]$ holds, where $\big(g^{i,\pmb{\alpha}^i}_\omega\big)'_-$ and $\big(g^{i,\pmb{\alpha}^i}_\omega\big)'_+$ are the left and right-hand derivatives of $g^{i,\pmb{\alpha}^i}_\omega$,  i.e.
\begin{equation*}
	\lambda\; \mathds{1}_{\big(-\infty, \; E^{i,\pmb{\alpha}^i}_T-A^i_T \big)}\big(\mathbf{Q}^i_T(\omega)\big) \leq  \mathbf{P}_T
	\leq \lambda\; \mathds{1}_{\big(-\infty, \; E^{i,\pmb{\alpha}^i}_T -A^i_T \big]}\big(\mathbf{Q}^i_T(\omega) \big).
\end{equation*}
As a result, $\mathbf{P}_T \in [0,\lambda]$.
Also using the zero clearing condition, we obtain 
\begin{equation}\label{eq:firstgen}
	\big\{\mathbf{P}_T \in (0,\lambda] \big\} \subseteq \cap_{i } \Big\{ \mathbf{Q}^i_T(\omega) \leq E^{i,\pmb{\alpha}^i}_T-A^i_T\Big\} \subseteq \big\{0 \leq   E^{\pmb{\alpha}}_T- A_T \big\}
\end{equation}
and 
\begin{equation}\label{eq:secondgen}
	\big\{\mathbf{P}_T \in [0,\lambda) \big\} \subseteq \cap_{i } \Big\{ \mathbf{Q}^i_T \geq E^{i,\pmb{\alpha}^i}_T-A^i_T \Big\} \subseteq \big\{ 0 \geq   E^{\pmb{\alpha}}_T- A_T \big\} .
\end{equation}
By~\eqref{eq:firstgen}, $\big\{E^{\pmb{\alpha}}_T- A_T <0 \big\} \subseteq  \big\{\mathbf{P}_T  =0 \big\}$. 
Likewise,~\eqref{eq:secondgen} yields $		\big\{E^{\pmb{\alpha}}_T- A_T >0 \big\} \subseteq  \big\{\mathbf{P}_T  =\lambda\big\}$.
This together with Assumption~\ref{ass:uniqueness} implies $\mathbf{P}_T  =\pmb{\xi}^{\pmb{\alpha}}$.
Hence $\mathbf{P}$ is the martingale closed by $\pmb{\xi}^{\pmb{\alpha}}$.
Applying Lemma~\ref{lem:terminalstrat} then yields the desired expression for the expected cost incurred by firm~$i$.

\section{Proof of Proposition~\ref{prop:equal_efoort}}\label{app:equal_efoort}

Lemma~\ref{lem:lemakkt} plays a key role in proving Proposition~\ref{prop:equal_efoort}. 
\begin{lemma}\label{lem:lemakkt}
    The first-order condition~\eqref{eq:foc} for the minimization problem~\eqref{eq:socialyoptimal} reduces to the following system of equations:
	\begin{align*}
		& \mathrm{k}^i \mu^i +\gamma^i (\mu^i)^2  \pmb{\alpha}^i (0)-\lambda \;\Phi\left(\frac{m(\pmb{\alpha})}{\norm{\mathrm{B}(T, \pmb{\alpha})}}\right) \mu^i = 0, \quad  i \in 1\,..\,n,\\
			& \mathrm{k}^i \mu^i +\gamma^i\big[ (\mu^i)^2+(\sigma^i)^2 \big] \pmb{\alpha}^i (t)-\lambda\; \Phi\left(\frac{m(\pmb{\alpha})}{\norm{\mathrm{B}(T, \pmb{\alpha})} }\right)\mu^i \\
			& \qquad \qquad -\; \frac{\lambda }{\norm{\mathrm{B}(T, \pmb{\alpha})} } \upphi\left(\frac{m(\pmb{\alpha})}{\norm{\mathrm{B}(T,\pmb{\alpha})}}\right) r^i(t, \pmb{\alpha}) =0, \quad i \in 1\,..\,n; \quad t\in 1\, .. \,  T-1, 
	\end{align*}
	where 
	\begin{equation*}
     	r^i(t, \pmb{\alpha}) = (\sigma^i)^2 (1-\rho^i) \big[1-\pmb{\alpha}^i (t)-a \big] + \sigma^i \sqrt{\rho^i} \; \sum_{j} \sigma^j \sqrt{\rho^j}\;\big[1-\pmb{\alpha}^j(t)-a \big],\;  t\in 1\,..\,T-1.
	\end{equation*}
\end{lemma}

\proof
In view of~\eqref{eq:g1}, we compute the gradient of the expected excess emissions:
\begin{align*}
	\nabla_\alpha  \mathrm{EE}(\pmb{\alpha}) &=  \Phi\left(\frac{m(\pmb{\alpha})}{\norm{\mathrm{B}(T,\pmb{\alpha})}}\right) \nabla_\alpha m(\pmb{\alpha})  + m(\pmb{\alpha}) \nabla_\alpha \Phi\left(\frac{m(\pmb{\alpha})}{\norm{\mathrm{B}(T,\pmb{\alpha})}}\right) \\
    	& \quad + \upphi\left(\frac{m(\pmb{\alpha})}{\norm{\mathrm{B}(T,\pmb{\alpha})}}\right) \nabla_\alpha \norm{\mathrm{B}(T,\pmb{\alpha})} + \norm{\mathrm{B}(T,\pmb{\alpha})} \nabla_\alpha \upphi\left(\frac{m(\pmb{\alpha})}{\norm{\mathrm{B}(T,\pmb{\alpha})}}\right).
\end{align*}
Using the identity $\upphi'(x) = -x\upphi(x)$ and the chain rule, we obtain
\begin{align*}
    \nabla_\alpha \upphi\left(\frac{m(\pmb{\alpha})}{\norm{\mathrm{B}(T,\pmb{\alpha})}}\right)
    &= -\frac{m(\pmb{\alpha})}{\norm{\mathrm{B}(T,\pmb{\alpha})}} \nabla_\alpha \Phi\left(\frac{m(\pmb{\alpha})}{\norm{\mathrm{B}(T,\pmb{\alpha})}}\right), \\
    \nabla_\alpha \norm{\mathrm{B}(T,\pmb{\alpha})}
    &= -\frac{1}{2\norm{\mathrm{B}(T, \pmb{\alpha})}} \nabla_\alpha \norm{\mathrm{B}(T,\pmb{\alpha})}^2.
\end{align*}
Substituting these into the expression for $\nabla_\alpha \mathrm{EE}(\pmb{\alpha})$ gives
\begin{align*}
     \nabla_\alpha  \mathrm{EE}(\pmb{\alpha}) 
    = \Phi\left(\frac{m(\pmb{\alpha})}{\norm{\mathrm{B}(T,\pmb{\alpha})}}\right) \nabla_\alpha m(\pmb{\alpha})
	- \frac{1}{2\norm{\mathrm{B}(T, \pmb{\alpha})}} \upphi\left(\frac{m(\pmb{\alpha})}{\norm{\mathrm{B}(T,\pmb{\alpha})}}\right) \nabla_\alpha \norm{\mathrm{B}(T,\pmb{\alpha})}^2.
\end{align*}
The computations of $\nabla_\alpha \mathrm{AC}(\pmb{\alpha})$, $\nabla_\alpha m (\pmb{\alpha})$ and $\nabla_\alpha \norm{\mathrm{B}(T,\pmb{\alpha})}^2$ are straightforward.
Setting $\nabla_\alpha R =\nabla_\alpha \mathrm{AC} + \lambda\nabla_\alpha \mathrm{EE}=0$  yields the required result.\ \finproof\\

\noindent
\emph{\textbf{Proof of Proposition~\ref{prop:equal_efoort}.}}
For each fixed firm $i$ and times $t,s \in 1\,..\,T-1$, subtracting the first order condition in Lemma~\ref{lem:lemakkt} at $s$ from that of at $t$ yields
\begin{equation*}
	\gamma^i \big[ (\mu^i)^2 + (\sigma^i)^2 \big] \delta^i - \frac{\lambda }{\norm{\mathrm{B}(T, \pmb{\alpha})} } \upphi\left(\frac{m(\pmb{\alpha})}{\norm{\mathrm{B}(T,\pmb{\alpha})}}\right) \left[r^i(t, \pmb{\alpha}) - r^i(s, \pmb{\alpha}) \right] = 0, 
\end{equation*}
where $ \delta^i = \pmb{\alpha}^i(t) - \pmb{\alpha}^i(s)$ and 
\begin{equation*}
	r^i(t, \pmb{\alpha}) - r^i(s,\pmb{\alpha}) = -(\sigma^i)^2 (1-\rho^i) \delta^i - \sigma^i \sqrt{\rho^i}\; S,  \quad \text{with} \quad  S=  \sum_j \sigma^j \sqrt{\rho^j} \; \delta^j.
\end{equation*}
Substituting this expression into the previous one yields
\begin{equation*}
	\gamma^i \big[ (\mu^i)^2 + (\sigma^i)^2 \big] \delta^i = -\frac{\lambda }{\norm{\mathrm{B}(T, \pmb{\alpha})} } \upphi\left(\frac{m(\pmb{\alpha})}{\norm{\mathrm{B}(T,\pmb{\alpha})}}\right)  \left[ (\sigma^i)^2 (1-\rho^i) \delta^i + \sigma^i \sqrt{\rho^i} S \right].
\end{equation*}
Solving for $\delta^i$, we obtain 
\begin{equation*}
	\delta^i = \frac{-S \lambda }{\norm{\mathrm{B}(T, \pmb{\alpha})} } \upphi\left(\frac{m(\pmb{\alpha})}{\norm{\mathrm{B}(T,\pmb{\alpha})}}\right)  \frac{\sigma^i \sqrt{\rho^i}}{M^i}
\end{equation*}
where 
\begin{equation*}
	M^i = \gamma^i \big[ (\mu^i)^2 + (\sigma^i)^2 \big] + \frac{\lambda }{\norm{\mathrm{B}(T, \pmb{\alpha})} } \upphi\left(\frac{m(\pmb{\alpha})}{\norm{\mathrm{B}(T,\pmb{\alpha})}}\right)  (\sigma^i)^2 (1-\rho^i)
\end{equation*}
Summing over $j$ gives:
\begin{align*}
    S &= \sum_j \sigma^j \sqrt{\rho^j} \; \delta^j 
		= \sum_j \sigma^j \sqrt{\rho^j} \left(\frac{-S \lambda }{\norm{\mathrm{B}(T, \pmb{\alpha})} } \upphi\left(\frac{m(\pmb{\alpha})}{\norm{\mathrm{B}(T,\pmb{\alpha})}}\right)  \frac{\sigma^j \sqrt{\rho^j}}{M^j} \right) \\
        	&  =\frac{-S \lambda }{\norm{\mathrm{B}(T, \pmb{\alpha})} } \upphi\left(\frac{m(\pmb{\alpha})}{\norm{\mathrm{B}(T,\pmb{\alpha})}}\right)  \sum_j \frac{(\sigma^j)^2 \rho^j}{M^j}.
\end{align*}
This yields a linear equation in $S$. 
Since the coefficient of $S$ on the right-hand side is positive, the only solution is $S = 0$.
Substituting back, we find $\delta^i = 0$ for all $i$, which concludes the proof.\ \finproof\\

\bibliographystyle{chicago}
\bibliography{biblio}

\begin{thebibliography}{}

\bibitem[\protect\citeauthoryear{Aïd and Biagini}{Aïd and Biagini}{2023}]{aid2023}
Aïd, R. and S.~Biagini (2023).
\newblock Optimal dynamic regulation of carbon emissions market.
\newblock {\em Mathematical Finance\/}~{\em 33\/}(1), 80--115.

\bibitem[\protect\citeauthoryear{Barrage and Nordhaus}{Barrage and Nordhaus}{2024}]{nordhaus2024}
Barrage, L. and W.~Nordhaus (2024).
\newblock Policies, projections, and the social cost of carbon: Results from the {DICE}-2023 model.
\newblock {\em Proceedings of the National Academy of Sciences\/}~{\em 121\/}(13), e2312030121.

\bibitem[\protect\citeauthoryear{Blanchard, Gollier, and Tirole}{Blanchard et~al.}{2023}]{blanchard2022fighting}
Blanchard, O., C.~Gollier, and J.~Tirole (2023).
\newblock The portfolio of economic policies needed to fight climate change.
\newblock {\em Annual Review of Economics\/}~{\em 15}, 689--722.

\bibitem[\protect\citeauthoryear{Carmona, Fehr, and Hinz}{Carmona et~al.}{2009}]{carmona2009}
Carmona, R., M.~Fehr, and J.~Hinz (2009).
\newblock Optimal stochastic control and carbon price formation.
\newblock {\em SIAM Journal on Control and Optimization\/}~{\em 48\/}(4), 2168--2190.

\bibitem[\protect\citeauthoryear{Carmona, Fehr, Hinz, and Porchet}{Carmona et~al.}{2010}]{carmona2010market}
Carmona, R., M.~Fehr, J.~Hinz, and A.~Porchet (2010).
\newblock Market design for emission trading schemes.
\newblock {\em SIAM Review\/}~{\em 52\/}(3), 403--452.

\bibitem[\protect\citeauthoryear{Guo, Tan, Gu, and Qu}{Guo et~al.}{2019}]{guo2019}
Guo, J.-X., X.~Tan, B.~Gu, and X.~Qu (2019).
\newblock The impacts of uncertainties on the carbon mitigation design: Perspective from abatement cost and emission rate.
\newblock {\em Journal of Cleaner Production\/}~{\em 232}, 213--223.

\bibitem[\protect\citeauthoryear{Heijmans}{Heijmans}{2023}]{Heijmans2023}
Heijmans, R.~J. (2023).
\newblock Adjustable emissions caps and the price of pollution.
\newblock {\em Journal of Environmental Economics and Management\/}~{\em 118}, 102793.

\bibitem[\protect\citeauthoryear{Hitzemann and Uhrig-Homburg}{Hitzemann and Uhrig-Homburg}{2018}]{hitzemann2018}
Hitzemann, S. and M.~Uhrig-Homburg (2018).
\newblock Equilibrium price dynamics of emission permits.
\newblock {\em Journal of Financial and Quantitative Analysis\/}~{\em 53\/}(4), 1653--1678.

\bibitem[\protect\citeauthoryear{Huang, Dong, and Jia}{Huang et~al.}{2022}]{huang2022}
Huang, Z., H.~Dong, and S.~Jia (2022).
\newblock Equilibrium pricing for carbon emission in response to the target of carbon emission peaking.
\newblock {\em Energy Economics\/}~{\em 112}, 106160.

\bibitem[\protect\citeauthoryear{{ICAP}}{{ICAP}}{2025}]{ICAP2025}
{ICAP} (2025).
\newblock Emissions trading worldwide: Status report 2025.
\newblock Retrieved on June 25, 2025 on \url{https://icapcarbonaction.com/en/publications/emissions-trading-worldwide-2025}.

\bibitem[\protect\citeauthoryear{Jiang and Yue}{Jiang and Yue}{2021}]{jiang2021}
Jiang, C. and Y.~Yue (2021).
\newblock Sensitivity analysis of key factors influencing carbon prices under the {EU ETS}.
\newblock {\em Polish Journal of Environmental Studies\/}~{\em 30\/}(4), 3645--3658.

\bibitem[\protect\citeauthoryear{{McKinsey Company}}{{McKinsey Company}}{2009}]{mckinesy2009}
{McKinsey Company} (2009).
\newblock Pathways to a low-carbon economy: Version 2 of the global greenhouse gas abatement cost curve.

\bibitem[\protect\citeauthoryear{Owen}{Owen}{1980}]{Owen1980}
Owen, D.~B. (1980).
\newblock A table of normal integrals.
\newblock {\em Communications in Statistics - Simulation and Computation\/}~{\em 9\/}(4), 389--419.

\bibitem[\protect\citeauthoryear{Rekker, Kesina, and Mulder}{Rekker et~al.}{2023}]{Rekker2023}
Rekker, L., M.~Kesina, and M.~Mulder (2023).
\newblock Carbon abatement in the {E}uropean chemical industry: assessing the feasibility of abatement technologies by estimating firm-level marginal abatement costs.
\newblock {\em Energy Economics\/}~{\em 126}, 106889.

\bibitem[\protect\citeauthoryear{Rockafellar}{Rockafellar}{1970}]{rockafellar1970}
Rockafellar, R.~T. (1970).
\newblock {\em Convex Analysis}.
\newblock Princeton University Press.

\bibitem[\protect\citeauthoryear{Seifert, Uhrig-Homburg, and Wagner}{Seifert et~al.}{2008}]{Seifert2008}
Seifert, J., M.~Uhrig-Homburg, and M.~Wagner (2008).
\newblock Dynamic behavior of {CO2} spot prices.
\newblock {\em Journal of Environmental Economics and Management\/}~{\em 56\/}(2), 180--194.

\bibitem[\protect\citeauthoryear{Smets and Wouters}{Smets and Wouters}{2007}]{smets2007shocks}
Smets, F. and R.~Wouters (2007).
\newblock Shocks and frictions in {US} business cycles: A {B}ayesian {DSGE} approach.
\newblock {\em American Economic Review\/}~{\em 97\/}(3), 586–606.

\bibitem[\protect\citeauthoryear{{United~Nations}}{{United~Nations}}{2015}]{united2015}
{United~Nations} (2015).
\newblock Adoption of the {P}aris agreement.
\newblock {\em Framework convention on climate change\/}.
\newblock 21st conference of the parties.

\bibitem[\protect\citeauthoryear{Woerdman, Roggenkamp, and Holwerda}{Woerdman et~al.}{2021}]{woerdman2015eu}
Woerdman, E., M.~Roggenkamp, and M.~Holwerda (2021).
\newblock {\em Essential {EU} Climate Law}.
\newblock Edward Elgar Publishing.

\bibitem[\protect\citeauthoryear{{World~Bank~Group}}{{World~Bank~Group}}{2025}]{Worldbank2025}
{World~Bank~Group} (2025).
\newblock State and trends of carbon pricing 2025.
\newblock Retrieved on June 25, 2025 on \url{https://openknowledge.worldbank.org/bitstreams/152de0c2-e2be-49d6-aec1-3be8ebad4f74/download}.

\bibitem[\protect\citeauthoryear{Wu, Wang, and Zhou}{Wu et~al.}{2023}]{Wu2023}
Wu, F., S.~Wang, and P.~Zhou (2023).
\newblock Marginal abatement cost of carbon dioxide emissions: The role of abatement options.
\newblock {\em European Journal of Operational Research\/}~{\em 310\/}(2), 891--901.

\bibitem[\protect\citeauthoryear{Yu and Mallory}{Yu and Mallory}{2020}]{Yu2020}
Yu, J. and M.~L. Mallory (2020).
\newblock Carbon price interaction between allocated permits and generated offsets.
\newblock {\em Operational Research\/}~{\em 20\/}(2), 671--700.

\bibitem[\protect\citeauthoryear{Zeidler}{Zeidler}{1986}]{zeidlernonlinear1}
Zeidler, E. (1986).
\newblock {\em Nonlinear functional analysis and its applications: I: Fixed-Point Theorems}.
\newblock Springer.

\end{thebibliography}

\end{document}